\documentclass[12pt,a4paper]{article}

\usepackage{amssymb,amsmath,amsfonts,color,eurosym,geometry,ulem,setspace,sectsty,comment,footmisc,caption,pdflscape,array}
\usepackage{listings}
\usepackage[utf8]{inputenc}
\usepackage[usenames,dvipsnames]{xcolor}
\usepackage[pdftex, hidelinks=true]{hyperref}
\usepackage{bookmark}
\usepackage{booktabs} % top/bottom rule
\usepackage{makecell} % make cell command
\usepackage{colortbl} % grey hlines in table
\usepackage[english]{babel}
\usepackage{float}
\usepackage[below]{placeins}
\usepackage{enumitem} % for itemlist
\usepackage{adjustbox} % adjusting table and figure size
\usepackage[authoryear]{natbib}
\usepackage{natbib}
\usepackage{verbatim} % comment out entire section
\usepackage[position=top]{subfig} 
\usepackage{threeparttable}
\usepackage{threeparttablex}
\usepackage[justification=centering]{caption}
\usepackage{graphicx}
\usepackage{wrapfig}
\usepackage{longtable}
\usepackage{hyperref}
\usepackage{afterpage} 
\hypersetup{
    colorlinks=true,
    urlcolor = blue,
    linkcolor = blue,
    citecolor   = black
}
\usepackage{tikz}
\usepackage{tabularx}
\usepackage{xcolor}
\usepackage{multirow}
\usepackage{multibib} % reference list for Appendix only
\newcites{appx}{Appendix References}
\usepackage[toc,page,header]{appendix}
\usepackage{minitoc} % table of contents for Appendix only
\usepackage{ragged2e}

\noptcrule 
\addto{\captionsenglish}{}
\usetikzlibrary{shapes,decorations,arrows,calc,arrows.meta,fit,positioning}
\tikzset{
    -Latex,node distance =1 cm and 1 cm,semithick,
    state/.style ={ellipse, draw, minimum width = 0.7 cm},
    point/.style = {circle, draw, inner sep=0.04cm,fill,node contents={}},
    bidirected/.style={Latex-Latex,dashed},
    el/.style = {inner sep=2pt, align=left, sloped}
}

\usepackage{CJKutf8}

\begin{document}

\doparttoc % Tell to minitoc to generate a toc for the parts
\faketableofcontents % Run a fake tableofcontents command for the partocs
% \part{}
% \parttoc % Insert the document TOC

\begin{titlepage}
%\title{The Unintended Consequences of Censoring Digital Technology -- Evidence from Italy's ChatGPT Ban\\\textit{PR: Alternative title: ``The Short-Term Effects of Generative AI on output/productivity: Evidence from Italy's ChatGPT Ban''}\thanks{We would like to thank Yashdeep Dahiya for outstanding research assistance.}
\title{The Heterogeneous Productivity Effects of Generative AI\thanks{We would like to thank Yashdeep Dahiya for outstanding research assistance and Samantha Eyler-Driscoll for manuscript editing and proofreading. This paper supersedes a previous version titled \href{https://arxiv.org/abs/2304.09339}{``The Unintended Consequences of Censoring Digital Technology -- Evidence from Italy's ChatGPT Ban''}, April 2023.}
\\[0.25cm]}

\author{
    David Kreitmeir\footnote{Department of Economics and SoDa Labs, Monash University (\href{https://davidkreitmeir.github.io/}{davidkreitmeir.github.io} and \href{mailto:david.kreitmeir1@monash.edu}{david.kreitmeir1@monash.edu})} \\  Paul A. Raschky\footnote{Department of Economics and SoDa Labs, Monash University (\href{https://praschky.GitHub.io/}{praschky.github.io} and \href{mailto:paul.raschky@monash.edu}{paul.raschky@monash.edu})}
    }

\maketitle
\vspace{-0.4cm}
\begin{abstract}
\noindent We analyse the individual productivity effects of Italy's ban on ChatGPT, a generative pretrained transformer chatbot. We compile data on the daily coding output quantity and quality of over 36,000 GitHub users in Italy and other European countries and combine these data with the sudden announcement of the ban in a difference-in-differences framework. Among the affected users in Italy, we find a short-term increase in output quantity and quality for less experienced users and a decrease in productivity on more routine tasks for experienced users. 

%Applying a synthetic control approach to daily Google search and Tor usage data shows that the ban led to a significant increase in the use of censorship bypassing tools. Our findings show that users swiftly implement strategies to bypass Internet restrictions but this adaptation activity creates short-term disruptions and hampers productivity.\\

\vspace{0.25cm}
\noindent \emph{JEL:} D8, J24, O33 \\
\noindent \emph{Keywords:} artificial intelligence, productivity \\

\bigskip
\end{abstract}
\setcounter{page}{0}
\thispagestyle{empty}
\end{titlepage}
\pagebreak \newpage

\begin{doublespace}

\section{Introduction}

%NOTES: 01/08/23
% Three contributions:
% 1. Productivity Effects of banning digital technology 2. Consequences of unilateral move in banning internet-based technology 3. Estimating productivity effects of a new general purpose technology using secondary data; building complement to existing lab and field experimental data.

% Stats on the importance of "coding" in the modern economy to show how important this sector is in the overall economy

%GitHub co-pilot

The public release of OpenAI’s ChatGPT provided near universal\footnote{Countries where ChatGPT is not accessible include  China, Eritrea, Iran, North Korea, Russia and Saudi Arabia, among others.} access to generative artificial intelligence (AI) tools at no or very low cost. Its subsequent quick adoption\footnote{According to OpenAI, by November 2023, ChatGPT was recording approximately 100 million weekly users. https://techcrunch.com/2023/11/06/openais-chatgpt-now-has-100-million-weekly-active-users/} broadened the discussion about the impact of generative AI on society and its potential to boost worker productivity by performing relatively complex tasks and producing (seemingly) novel output, all while requiring only minimal technological knowledge on the part of users. However, ChatGPT also has the tendency to produce wrong or faulty outputs (e.g., “hallucinations”) that, in the absence of expert knowledge, are difficult to detect and costly to rectify and might ultimately undermine the productivity of some workers \citep{dellacqua2023}.

One of the focal points in the discussion on generative AI's societal impact is its ability to create and generate new content and knowledge. Similarly to prior advances in AI, generative AI can enhance productivity by replacing more routine tasks \citep{brynjolfsson23,Kanazawa2022,noy23,peng2023} or improving users' decision accuracy \citep{Kleinberg17,almog2024ai,cho2023baseball}. The feature that sets this generation of AI apart from previous ones is that, with its access to the universe of online knowledge, generative AI combines domain-specific information with rules, lending it the ability to create new content and ultimately opening the possibility of extending the production possibility frontier beyond an individual’s current level of training or expertise. 

% DK TO PR: I think here would be a good point to sagway to coding vs. producing text (the most common application) and then using this great study?
However, the accuracy of current generative AI models' performance in some tasks such as text summarization or generation, combined with its clarity and confidence of delivery, might create the \emph{illusion} that it enhances productivity in other domains. Inaccurate, faulty or ``hallucinated'' output may not be immediately detected and could be used as an input in a knowledge worker's subsequent production flow. For instance, \cite{kabir2024} analysed ChatGPT's response to 517 programming questions and found that 52\% of its answers were incorrect and that the users presented with these answers overlooked these errors 39\% of the time. Nevertheless, users still tend to prefer to use ChatGPT because of its comprehensive responses \citep{kabir2024} and the confident language of the responses \citep{li2023}. The biggest online discussion and help forum for developers, stack overflow, has banned the use of LLM generated content on the forum because the rate of ``getting correct answers'' from these tools, it too low.\footnote{https://meta.stackoverflow.com/questions/421831/policy-generative-ai-e-g-chatgpt-is-banned}

For some tasks (e.g., content writing), the process of detecting faulty output or rectifying generative AI--driven errors might be quick, while for others (e.g., software development), the same process can be tedious and time consuming.\footnote{While the motivation for the relatively early public release of ChatGPT and other large-language models (LLMs) was to improve their performance with the human-generated data collected from user interactions with the tools, their output quality remains noisy, and expert knowledge is often required to accurately judge this quality. Moreover, \cite{chen2023} show that ChatGPT's performance on a number of tasks, including generating code, actually declined from version 3.5 to version 4.0, calling into question whether its performance and accuracy will continuously improve over time. In addition, \cite{delriochanona2023large} show that the widespread use of ChatGPT has led to a decline in usage of online help forums such as Stack Overflow, which in return will decrease the human-generated ground truth data that can be used to improve AI models.} There are also wide differences in the accuracy of generative AI output, driven not only by the complexity of the underlying task but also the size and quality of the underlying training data. The tools can leverage a very large online text corpus to predict the next word in  tasks such as creative writing and chatting, but the training data for software development and code creation are limited to a relatively small number of online forums (e.g., Stack Overflow), where the ground truth can be noisy.\footnote{For example, for less routine, more complex and more niche questions, the answers provided on Stack Overflow are not necessarily correct or are just initial solution suggestions instead of working solutions. While these suggestions might have received upvotes, signaling to the LLM their ``usefulness'' for its training, in reality, the content of the answers and ChatGPT subsequent output might be only an untested and ultimately not functioning code routines.}

In cases where the underlying task is more complex and the output requires accuracy to be ultimately useful (e.g. software development), relying on generative AI might prolong task completion and decrease workers' output quality \citep{dellacqua2023}. Such problems might be more acute among less experienced workers, who may have less domain knowledge and require more time to detect and correct errors. Less experienced workers might also be more prone to continue using the tool because the alternatives (e.g., acquiring the knowledge and skills themselves) appear to be even costlier.

In this paper, we use observational data to analyse the heterogeneous effects of ChatGPT on the output quantity and quality of experienced and less experienced software developers.  In particular, we exploit Italy's sudden announcement of a ChatGPT ban as a natural experiment to examine the ban's short-run effects on GitHub users' productivity. We find that the ban had no systematic effect on the overall output of more experienced developers and only some small negative effects on their output for more routine tasks (resolving issues and debugging). However, among less experienced users, the short-term lack of access to ChatGPT increased both the amount of output and its quality. For this group of users, the likelihood that we observe any output-related activity on GitHub is approximately 10\% higher for the two business days following the ban. This effect size shrinks for the subsequent days. In the same vein, we find some tentative evidence that Internet users in Italy adapted fairly quickly to the legislation by increasing their use of virtual private networks (VPNs) and encrypted routing to circumvent the ban. 

Finding high-frequency and consistently measured, observational data  on the output of knowledge workers that is comparable across countries is challenging.  We follow a the approach from a number of recent papers \citep[e.g.,][]{mcdermott21,holub2023,shen2023} that have already exploited the temporal granularity of GitHub activity data and used this data as a proxy for software developers' productivity. When using this data to proxy productivity, we rely on a number of assumptions which we test in a subsequent robustness section. We show that the results are not driven by changes in working hours or the level of complexity of the task. Considering that the ban was implemented close to a major holiday in the treatment and control countries, we also show that the changes in behaviour are not driven by seasonal factors. Our  results are also robust to the use of a set of alternative outcome variables and also analysing the effect at the user--repository--day level.

Our results present some first, nonexperimental empirical evidence on the effects of restricting access to generative AI on workers' performance in more complex tasks. Importantly, we show that the effects of generative AI are heterogeneous by worker's experience. 

 %It is likely that the short-term drop in output is a result of users' adaptation behaviour and their time spent to circumvent the ban rather than on productive activities. Overall, our findings reveal that potentially well intended restriction of Internet access, for the purpose of protecting user data and privacy, has unintended, negative consequences on productivity that are short-lived.  

Our study complements the existing, largely experimental, literature on the effects of generative AI on worker productivity in less complex tasks (e.g., content writing, customer support), where generative AI output is less error-prone, by examining a setting with more complex tasks, where AI-generated output can be less accurate or more faulty. Existing work by \cite{brynjolfsson23} and \cite{noy23} has found mainly positive productivity effects of generative AI and stronger effects for less experienced workers in the contexts of customer support and content writing tasks. In contrast, our results suggest that, for more complex tasks (e.g., code development),\footnote{Related work by \cite{peng2023} and \cite{chatterjee2024impact} has found positive effects of GitHub Copilot on both the productivity and job satisfaction of software developers. While GitHub Copilot is also an AI-based tool developed by GitHub and OpenAI, it is specifically a code completion tool. In contrast, ChatGPT is a chat-based (rather than auto-complete) tool and can be used to produce entirely new code segments/programs based on a human language prompt; in such cases, the accuracy of generative AI is highly variable \citep{kabir2024}.} generative AI does not necessarily boost the productivity of less experienced workers and can even decrease their output quantity and quality. Our results, therefore, confirm the findings of \cite{dellacqua2023} in a controlled experiment environment that, for tasks beyond ChatGPT's current capabilities, using ChatGPT increases the time a worker spends on a task. We complement their results by highlighting the differential effects by knowledge worker's level of experience. Our finding of heterogeneous effects for more complex tasks also empirically complements the larger discussion in economics on technological change and inequality in the labour market \citep[e.g.,][]{Acemoglu2002,autor2003,Goldfarb2019,Acemoglu2020}, in particular the productivity and labour market effects of AI  \citep[e.g.,][]{brynjolfsson2017,agrawal2019,Acemoglu2021,eloundou2023}. 

%Our paper also speaks to the large literature on the effects of restricting access to the Internet and cloud based technologies on society \citep[][e.g.,]{roberts2018,hobbs2018} documents how the Chinese Firewall adversely affects Internet speed in China while \cite{hobbs2018} shows how Chinese users react to the sudden ban of Instagram by an increased used of  circumvention tools. \citep{chen2019} conduct a field experiment where they randomly provide Chinese Internet users with access to censorship circumvention tools. They find that access to uncensored Internet in itself does not increase the demand for less censored online information. However, in combination with a treatment that encourages the use of uncensored Internet, and the associated consumption of less censored media leads to a change in the level of information and political beliefs. %In some cases, the blocking of particular websites is also used to counter foreign propaganda and misinformation, such as the ban of the Russian social media platform VKontakte in Ukraine \citep{golovchenko2022}. We complement this literature by focusing on censorship in a Western, democratic country and evaluating its effects on individual productivity of knowledge workers.

The paper is organised as follows: Section 2 provides some background on ChatGPT and the Italian ban on the technology in 2023. Section 3 describes the data. Section 4 presents empirical results on the ban's effect on worker productivity, and Section 5 concludes.

\section{ChatGPT and the Italian Ban}

ChatGPT, an LLM created by US startup OpenAI, has been used by millions of people since it launched in November 2022. Trained on a vast corpus of text data from the Internet as it was in 2021, this large-scale AI language model uses a transformer-based neural network to process natural language. During the training process, the model learned to identify patterns and relationships between words, phrases, and sentences, enabling it to generate text.%\footnote{Such as this very paragraph.} 

ChatGPT is accessible via a public website (chatgpt.openai.com) or an application programming interface (API), and almost anyone\footnote{Before Italy, countries including China, Russia and North Korea had already banned ChatGPT.} can sign up for a free account. The interface is designed like a chat environment where the user writes ``prompts'' and ChatGPT answers. Interactions can range from casual chats and search-like queries to more complex exchanges such as creative writing of a text or creation of recipes based on prompts. ChatGPT can also write code in multiple programming languages on the basis of a simple prompt.

On April 1, 2023, the Italian data protection authority (Garante per la protezione dei dati personali) blocked use of the ChatGPT chatbot, citing privacy concerns, and announced an investigation into OpenAI's compliance with the European Union's General Data Protection Regulation (GDPR). In particular, the authority stated that there was no legal basis for the mass collection and storage of personal data to train the algorithms underlying the platform's operation.\footnote{Shiona McCallum, ``ChatGPT banned in Italy over privacy concerns'', BBC 01/04/2023, https://www.bbc.com/news/technology-65139406} There are a number of reasons to expect that the ban did not have an effect on software developers. During the ban access to other Natural Language Processing (NLP) powered tools that are powered by Open AI, such as GitHub CoPilot, was not affected. In contrast to ChatGPT which was trained on the universe of available online text corpus, GitHub Copilot was specifically trained only on code-repositories and build to produce context-aware codes. It provides more accurate results and is also widely used by code-developers. It is important to note, that our study only estimates the effects of a ban of ChatGPT on developer's output and not a ban on all LLM based coding support tools. Alternative generative AI tools such as Claude and Bard which were launched a few weeks prior to the ban were still accessible. There were also multiple ways for programmers to immediately circumvent the ban (i.e. VPN) and guides on how to do that where shared online.\footnote{https://www.programmareinpython.it/blog/chatgpt-bloccato-in-italia-che-fare/}

However, anecdotal evidence shows that shortly after the inception of the ban, Italian developers went online and complained about the disruption caused by restricting access to ChatGPT, ``[...] a tool that has become an essential part of [their] daily routine'' as software developers.\footnote{Semero, Anto ``ChatGPT Banned in Italy: Mamma Mia! What’s Going On?'' https://medium.com/@antonellosemeraro/chatgpt-banned-in-italy-mamma-mia-whats-going-on-97c44284e331, April 2 2023.} Concerns were raised that restricting access to ChatGPT during a time when the field of software development is increasingly fast-paced, poses a severe threat to the competitiveness Italian developers and businesses.\footnote{https://www.programmareinpython.it/blog/chatgpt-bloccato-in-italia-che-fare/} Whether the ban was effective or not is ultimately an empirical question, which this paper aims to shed light on.

The ban was lifted in late April after OpenAI responded to the data protection authority's privacy concerns.\footnote{Shiona McCallum, ``ChatGPT accessible again in Italy'', BBC 28/04/2023, https://www.bbc.com/news/technology-65431914}

%A priori it is unclear how the ban might affect individual productivity. As shown by recent papers, ChatGPT can enhance productivity of coders \citep[e.g.,][]{kashefi2023,sobania2023}. Therefore, the sudden ban can have negative effects on individual productivity. However, ChatGPT is still a fairly young technology and there is no reliable data available on its diffusion within the economy. The media hype around the technology might paint a misleading picture of the  adoption by companies and individual knowledge workers. As such, the ban might not have any systematic impact on individual productivity. It is also possible, that 

\section{Data}

\paragraph{GitHub Data} GitHub is the world’s largest online code hosting platform, used for storage of and joint work on coding projects (so-called repositories).\footnote{The programming languages most commonly represented in GitHub repositories are \texttt{Python} (17.38\%), \texttt{Java} (11.77\%), \texttt{Go} (10\%), \texttt{JavaScript} (9.95\%),  and \texttt{C++} (9.66\%). In comparison,  \texttt{R}-related repositories account for only 0.074\% of all pull requests on GitHub. https://madnight.github.io/githut } All modifications to a GitHub repository are automatically timestamped and stored, and GitHub permits tracking of any iterations of specific files and lines of code. Every action taken by a team member is automatically recorded, with details about the kind and substance of the modification, the files and code lines affected, and the date the changes were performed. Anyone with access to a repository can examine and download the history of iterations and actions, and given GitHub's history of developing open-source software, a significant portion of its repositories are not access restricted, meaning that the project activity information is available to everyone. Thus, public GitHub repositories provide a direct, real-time measure of labour activity for millions of software and code developers worldwide \citep{mcdermott21}.\footnote{GitHub data have been used in empirical research on software developers' productivity during the onset of COVID-19 \citep{forsgren2021}, the impact of COVID-19 on daily and weekly patterns of individual labour allocation \citep{mcdermott21}, the effects of working from home on individual productivity \citep{shen2023}, the effect of air pollution on individual output \citep{holub2023}, and the relationship between social links and the likelihood of joining professional software development teams \citep{casalnuovo2015}.} 

We access individual-level, real-time activity data from the \href{https://www.gharchive.org/}{GitHub Archive} for GitHub users in Italy (treatment) and Austria, France, and Spain (control) in the week prior to and that immediately after the ChatGPT ban in Italy (March 27--April 9, 2023).\footnote{We chose these three countries because all of them are part of the European Union and share a common land or sea border. In settings like ours, it is difficult to find objective criteria that help guiding the choice of comparable units. To provide further confidence that our results are not based on the choice of control countries, we conduct a leave-one-out analysis in Figure \ref{fig:leave-one-out} and show that the results are not driven by the composition of the control group.} To account for the Easter break starting with Good Friday on 7 April---a public holiday in all four sample countries--- we restrict the post-treatment period to Monday 3 April to Thursday 6 April and the corresponding pre-period to Monday 27 March to Thursday 30 March.\footnote{For more details on the study's time line and a graphical illustration please see \ref{sec:time-line} in the appendix.} GitHub Archive is hosted on Google’s BigQuery warehouse system and contains all public event data of GitHub user, which is updated daily and can be accessed with a query on Google's cloud infrastructure. GitHub user information such as the year of GitHub user account creation was downloaded with the GitHub GraphQL API.\footnote{For more information on how we retrieve GitHub user location information, please see \ref{sec:github-user} in the appendix.} The two datasets are merged via the unique GitHub user login.

We use the individual-level action data to construct two sets of baseline outcome variables: The first group captures \emph{output quantity and quality} and includes aggregate \emph{Output} limited to ``productive'' actions, aggregate \emph{Output} as defined by \cite{shen2023}, aggregate \emph{Output} as defined by \cite{holub2023}, and the \emph{Pull request (PR) merge ratio} quantifying how many of a user's suggested code edits were accepted by the repository (project) owners. The second group gauges \emph{task choice and complexity} and comprises \emph{PR opened}, \emph{Avg. lines added per opened PR}, \emph{Avg. lines added per merged PR}, \emph{Easy issue closed}, and \emph{Interactive activity}. A detailed description of the construction and definition of all outcome variables is provided in appendix Table \ref{tab:var-defs}.
%Our preferred measure for GitHub users' output are \textit{Release} events. A release is a deployable version of a software project that make it available for a wider audience and easy for users to download and install the software. It typically includes a set of release notes, documentation, and binaries or source code.\footnote{See https://docs.GitHub.com/en/repositories/releasing-projects-on-GitHub/about-releases}. 

On a daily basis, GitHub actions are relatively rare events at the user level. Hence, we transform each count variable into a binary indicator that equals 1 if one of the actions in a category is recorded for the user on a given day, and 0 otherwise.\footnote{The distribution of day--user-level counts of the main event variables during the sample period is presented in Figure \ref{fig:distribution-counts} in the appendix.} Descriptive statistics at the user--day and user level are presented in appendix Table \ref{tab:summary-user}. To use GitHub actions as a measure for developers' productivity, we need to ensure that the ban did not affect developers' working hours. While it is not possible to have data on exact working hours for developers, we are still able to check if time spent on GitHub activity during the day has changed between the pre and post period. Using the timestamp data for each individual GitHub action, we are able to show that the distribution of GitHub activities across hours of the day did not change over the period (see Figure \ref{fig:distribution-work-activity-Italy}).

% We also use the (log of) number of total GitHub events by a user, $Events$, the sum of Push and Pull events, \textit{PushPull} and an aggregate \textit{Output} measure based on  \citet{holub2023}, which is the sum of Push, Pull Requests, Pull Request Comments, Commit Comments, Create, and Issues. In addition, we use information about the users organisational affiliation to distinguish between GitHub users who have entered an organisational affiliation (i.e. company, research organisation) and those without an organisational affiliation.

\paragraph{Package Repositories} We compile a list of packages hosted on GitHub for ten analytical programming languages: \texttt{C}, \texttt{C++}, \texttt{Go}, \texttt{Java}, \texttt{JavaScript}, \texttt{Julia}, \texttt{Perl}, \texttt{Python}, \texttt{R}, and \texttt{Rust}. We rely in the first instance on the community--curated \href{https://GitHub.com/sindresorhus/awesome?tab=readme-ov-file}{``Awesome Lists''} to locate GitHub repositories for ``popular'' packages in each language. In a second step, we scrape the information on all packages hosted on the official software repositories for \texttt{Python} (\texttt{\href{https://pypi.org/}{pypi}}), \texttt{R} (\texttt{\href{https://cran.r-project.org/}{CRAN}}) and \texttt{Julia} (\texttt{\href{https://GitHub.com/JuliaRegistries/General}{JuliaRegistries}}) to retrieve information on each package's GitHub repository. We make use of the standardized GitHub URL structure to identify the \textit{owner} and \textit{name} of a package repository.\footnote{The stylized URL for a package repository is \texttt{https://GitHub.com/[account name hosting the repository]/[repository name]} (for instance, \url{https://github.com/numpy/numpy}).} To identify the GitHub user accounts other than the owner that contribute to a package repository, we use information on each individual GitHub user’s activity from January 2011 until March 2023. We restrict the list of \textit{GitHub event types} to ``productive'' events to select primarily accounts that made at least one substantial contribution to a package repository.\footnote{Our set of ``productive'' event types comprises \texttt{PullRequestEvent}, \texttt{PullRequestReviewEvent}, \texttt{PullRequestReviewCommentEvent}, \texttt{PushEvent}, and \texttt{ReleaseEvent}.} Moreover, we winsorise the sample at the 1st and 99th percentiles to safeguard against outliers.\footnote{Note that we exclude \texttt{bot} accounts from the list of contributors prior to winsorising.} % and that we do \textit{not} a priori exclude accounts that are the owner of a package but in the 1st percentile of contributions. 
Our final list of package contributors and owners comprises $483,855$ unique GitHub user accounts, of which $5,916$ are part of our baseline sample.

\section{Effect of the ChatGPT Ban on GitHub Output}

To analyse the effect of the Italian ChatGPT ban on GitHub users' output, we estimate variants of the following difference-in-difference (DID) specification: 
\begin{align}
%\begin{split}
    Y_{it} =& \beta D_{it} + \alpha_{i} + \lambda_{t} + \gamma_{dow} + \sigma_{d} \times t + \epsilon_{it}, \label{eq:did} 
%    \end{split}
\end{align}
where $Y_{it}$ denotes the outcome variable, i.e. one of the user-specific output and task variables, and $D_{it}$ is the treatment indicator variable equaling 1 for Italian users in the first four work days post the ChatGPT ban. The parameters of interest is $\beta$. Time-invariant differences between users, including ability and experience, are captured by user fixed effects $\alpha_{i}$, while day (date) fixed effects $\lambda_{t}$ account for daily fluctuations in coding output across users. Additionally, we account for differences in working behaviour across week-days via day-of-the week $\gamma_{dow}$ and include control treatment and control group specific time trends to safeguard against differential time-trends in coding activity between Italian Github users and their European peers.\footnote{We estimate alternative specifications of our baseline regression model. Including country-specific linear time-trends or excluding time-trends altogether leaves our baseline estimates quantitatively and qualitatively stable (Table \ref{tab:did-wo-trend} and Table \ref{tab:did-ctry-trend} in the appendix).}

To check for potential pre-trends and investigate how the estimated effect evolves over time, rather than averaging over the whole window as in the generalized DID specification, we also estimate the following event-study specification:
\begin{align}
%\begin{split}
    Y_{it} =& \sum_{\tau = -4}^{-2} \beta_{\tau}  D_{it}^{\tau} + \sum_{\tau = 0}^{3} \beta_{\tau} D_{it}^{\tau} + \alpha_{i} + \lambda_{t} + \gamma_{dow}  + \sigma_{d} \times t  + \epsilon_{it}, \label{eq:event-study}
%    \end{split}
\end{align}
where $D$ is a dummy variable equalling one for observations in the treatment group at event--day $\tau$ and zero otherwise; with $\tau = -1$ serving as the (excluded) reference period. In both specifications, we cluster standard errors at the user level.
% DAVID: habe die trends hier rein in die equationn auch wenn sie gedropped werden aufgrund von multicllinearity aber wollte nicht dass sich einer darauf aufhaengt oder was meinst du?

\subsection{Baseline Results}

Table \ref{tab:did-main} presents our baseline DID model estimates for the coefficient of interest $\widehat{\beta}$. The upper panel reports the average treatment effect of the ChatGPT ban on output quantity and quality for four different samples (overall, less experienced, experienced and package contributors), while Panel B presents baseline estimates for task choice and complexity. We find that the ban has \textit{overall} no significant effect on the output quantity and quality of Italian users or their choice of tasks with the notable exception of a negative and significant impact on their ability to close issues. These aggregate results, however, disguise important heterogeneity across users.  

% Talk here about heterogenous effects on output quantity and quality

% Here on task choice/complexity
In Panel B, we provide suggestive evidence that the estimated effect on output quantity and quality of \textit{less experienced} users is not result of a change in tasks. In particular, we do not see that the number of pull requests send for review (column 1) changes nor their complexity (column 2). Further, restricting the focus to lines added on pull requests that were actually merged (column 3), reveals a significant positive effect. While this increase could reflect less efficient code by Italian users post-ban, our estimates provide no support for the notion that the increase in the PR merge ratio is the result of decrease in merged PR task complexity. Similarly, we do not observe a switch to  easier issues (column 4) nor to collaborative engagement  instead of coding activity (column 5).\footnote{In unreported results, we test if the estimated effect on \textit{easy} issues is the result of restricting the attention to the resolution of those issues. If we consider any form of engagement with easy issues---i.e. commenting, opening or closing---we find that the probability of engaging with easy issues declines for less experienced users (significant at the 10\% level), while no change is observed for more sophisticated users.}  For \textit{experienced} users, we find null effects of the ban on task allocation and complexity, while the ban appears to have some effects on \textit{package contributors}. In particular, the decline in the probability of opening a pull request and resolving easy issues might indicate that package contributors tackle less but harder tasks after the ban.

% Talk about event study results here: Focus on Figure for experienced users ()

\begin{landscape}
\begin{figure}[H]
\centering 
\caption{Less Experienced Users -- Event-Study Estimates}
\label{fig:event-study-inexp}
\includegraphics[width=1.5\textwidth]{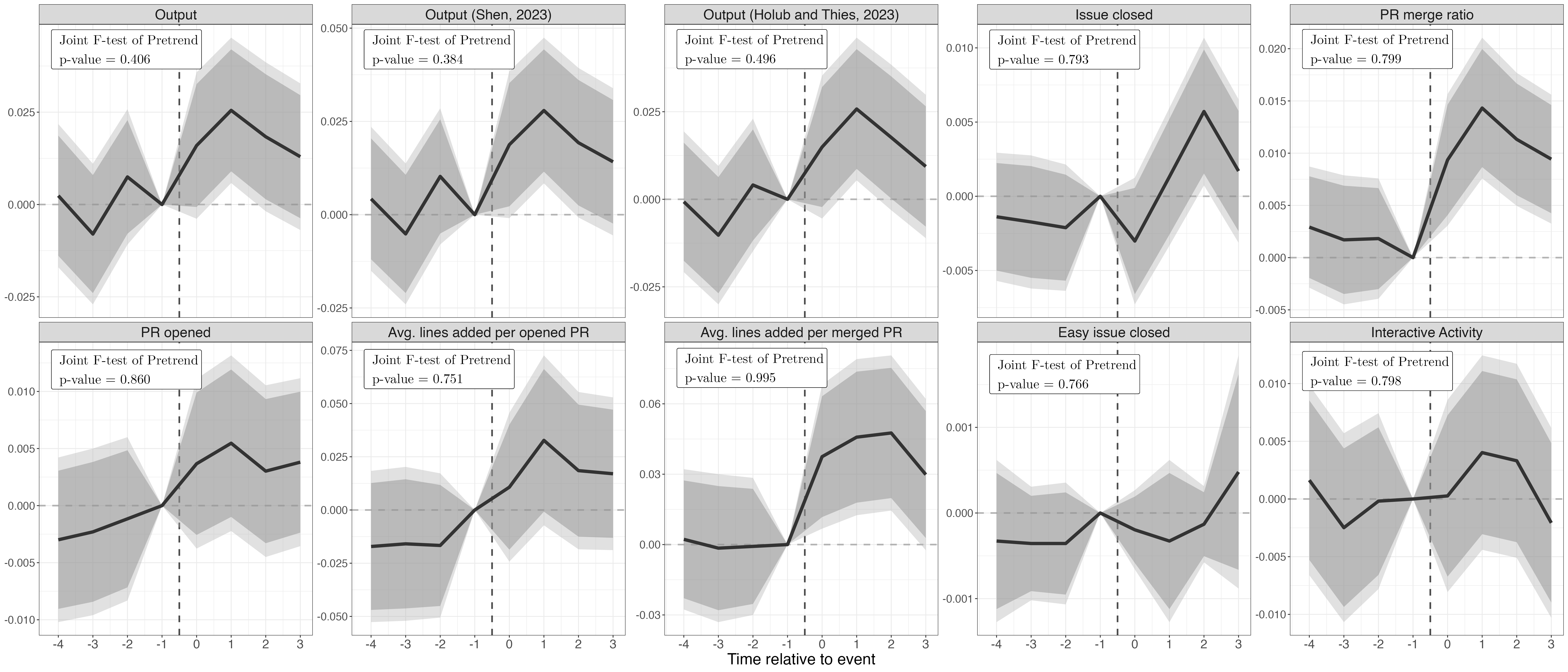}
\captionsetup{justification=justified, singlelinecheck=off} 
\caption*{\footnotesize \textit{Notes:} Event-study estimates across outcomes for ``less experienced'' GitHub user accounts (created after or in 2017). The sample period spans March 27--30 (\textit{Pre}) and April 3--6 (\textit{Post}). All specifications include user, time, and day-of-the-week fixed effects. A log plus one transformation is applied to \textit{Avg. lines added per PR} (opened or merged).  95\% (90\%) confidence intervals for robust standard errors clustered at the user level are depicted in light (dark) grey.}
\end{figure}
\end{landscape}

Figure \ref{fig:event-study-inexp} presents the estimated event-study coefficients $\beta_{\tau}$ from specification \ref{eq:event-study} for \textit{less experienced} users. Importantly, a joint F-test of whether all coefficients prior to the ban are jointly zero cannot be rejected at conventional levels for any outcome, alleviating concerns about preexisting trends. Further, the event-study results reveal that the treatment effect for less experienced users peaked two days after the ban. Figure \ref{fig:event-study-exp} and \ref{fig:event-study-pkg} in the appendix present the corresponding event-study estimates for \textit{experienced} users and \textit{package contributors}, respectively. Results for this subset of users reveal that point estimates are rather volatile over time, particularly for \textit{package contributors}, and reveal no clear pattern except for \textit{issue closed}. Overall, our event-study results provide support for our main DID findings that the ban actually increased the productivity of \textit{less experienced} coders, while more sophisticated users were largely unaffected.

\subsection{Robustness Checks}

We conduct a number of robustness checks and present the results in the appendix. 

\paragraph{Alternative outcomes:} First, we consider a number of additional outcomes in Panel A of Table \ref{tab:did-alt-out}. We show that our results are largely robust to alternative measures for output quantity (\emph{Any Event} and \emph{Commit}), output quality (\emph{PR merged} and \emph{PR merge ratio \citep{holub2023}}) as well as task complexity (\emph{Avg. files edited per merged PR}). In Panel B, we additionally show that our baseline findings are qualitatively stable at the intensive margin when we use continuous outcome variables instead of binary indicators.
%\footnote{Table \ref{tab:did-alt-out} also presents estimates on \texttt{Releases}, confirming the results presented in a previous version of the paper.}
\paragraph{Placebo tests:} Second, the overall changes in output quantity and quality could be driven by unobserved factors occurring at the same time as the introduction of the ban or general differences in Italian Github user activity relative to their European peers in the work week leading up to the Easter break. To address these concerns, we undertake two placebo exercises. In the first placebo test, we assume that the ban was implemented on the weekend prior to our \textit{actual} pre-treatment period from Monday 27 to Thursday 30 March 2023. The estimated \textit{placebo} treatment effects are presented in Table \ref{tab:did-placebo-2023} and are relatively small in magnitude and vastly insignificant. Importantly, we don't find the heterogeneous effects across users uncovered in our baseline analysis. Our second \textit{placebo} treatment period spans the work week prior to Good Friday on 15 April 2022. Reassuringly, our placebo estimates (Table \ref{tab:did-placebo-2022}) are again exceedingly insignificant and do not follow a clear and comparable pattern to our baseline results.

\paragraph{Repository-level Analysis:} For a subset of users working on multiple repositories, we construct a new panel dataset at the user--repository--day level (Section \ref{sec:repo-results} in the appendix).\footnote{Note that the set of public repositories is also restricted to those with at least one user from both the control and treatment group.} This allows us to include repository (project) and user fixed effects. Exploiting only within user-repository variation projects we estimate the effect of the ban on users while holding the complexity of a project constant. The results in Table \ref{tab:did-repo-main} largely confirm the patterns from our baseline, user-level, analysis. For less experienced users, we still observe a positive and significant effect on output quantity and quality, while the likelihood of closing an issue for is still negatively affected by the ban. Interestingly, we now also find some suggestive evidence for the negative impact of the ban on experienced users' output quantity relative to that of their European peers working on the same repository. Moreover, we find additional support for the negative effect of the ban on experienced users when conditioning on programming language and restrict our attention to highly complex official package repositories (Table \ref{tab:did-repo-lang}).\footnote{Because of the limited number of observations for package repositories of other programming languages, our analysis is, unfortunately, constrained to official \texttt{python} packages from \texttt{PyPi}. In addition, we exclusively present results for output quantity and quality in \ref{tab:did-repo-lang} due to the limited variation in the data for most outcomes concerning task choice and complexity.} 

\paragraph{Additional Robustness Checks} We conduct a number of additional sensitivity checks. % First, we extend the pre-period to additionally comprise the period from Monday 20 to Thursday 23 March 2023.    
First, we conduct a "Leave-One-Out" analysis to further alleviate concerns that our baseline findings are dependent on our selection of control group countries. In particular, we show in Figure \ref{fig:leave-one-out} that our results are robust to excluding users from each of the control group countries. Finally, we address concerns that our results could suffer from an overrejection of the null hypotheses as a result of reuse of the identifying exogenous variation for multiple outcomes (and subsamples). Table \ref{tab:did-rwolf} reports p-values (in brackets) that are corrected for multiple hypothesis testing following the procedure described in \citet{Romano_Wolf_2005_JASA, Romano_Wolf_2005_Econometrica, Romano_Wolf_2016}.\footnote{In recent work, \cite{heath2022} show that the employed \textit{Romano--Wolf} correction procedure performs well in a multitude of settings and across different dimensions.} Importantly, output quantity as defined in \cite{shen2023} and, in particular, output quality retain their significance even after correcting for multiple hypothesis testing.

% Third, in Table \ref{tab:did-log}, we replace the binary outcome variables with continuous variables and show that our results are also robust on the intensive margin. % Fourth, while we believe that the unexpected implementation of the ban in only one European country makes the assumption of parallel trends credible in our empirical setting, we nevertheless conduct a sensitivity analysis of possible post-treatment violations of parallel trends relative to pre-treatment violations. Appendix Figure \ref{fig:honestdid-magnitudes} presents relative magnitudes bounds, $\Delta^{RM}\left( \Bar{M} \right)$, for the average treatment effect over the first two post-treatment days and main outcomes under different values of $\Bar{M}$, as suggested by \cite{honestdid}.\footnote{For instance, $\Bar{M} = 1$ restricts the post-treatment violations of parallel trends over the first two post-treatment days to be no larger than the maximal pre-treatment violation of parallel trends over two consecutive days.} 

\subsection{Discussion}

Data from public GitHub repositories include code and software projects from a variety of organisations and individuals. Some of these are open-source development projects (e.g., APIs) from private-sector companies, some are general open-source projects developed by a community of volunteers (and therefore are closer in character to public goods), and others are owned by research organisations or individual developers. Given data limitations, it is not possible to distinguish the type of project.

It is possible that some Italian users immediately used tools (e.g., VPNs) to circumvent the ban. Using data on Google searches for VPN services and usage data for TOR\footnote{The TOR (The Onion Router) network is an open-source overlay network of thousands of network relays that conceals a user's IP address. Unfortunately, we cannot access actual VPN usage data at daily level.} (see appendix \ref{sec:google-tor}), we show a sudden jump in circumvention activity among Italian Internet users in the days after the ban. Despite the easy access to circumvention technology, many corporations and organisations actually prohibit the use of VPN and TOR tools on their devices and networks, implying that their use may be limited to mainly private devices and home networks. More importantly, we still find systematic effects on output despite this circumvention activity, and one can interpret our results as a lower bound. Another concern is that our finding of heterogeneity between less experienced and experienced users could be driven by the latter's greater skill in circumventing the ban. However, the systematic effects of the ban on tasks related to closing issues and the negative output effect detected in repository-level analysis suggests that this is not the case. 

There are a number of follow-up questions that we are unable to empirically analyse because of limitations in the data. First, while generative AI might disrupt the production flow of less experienced workers by providing faulty results, another possibility is that ChatGPT is simply a distraction. While it is not clear how this would explain the effect heterogeneity, more detailed data on the actual use of ChatGPT could help inform the design of workplace policies around generative AI.\footnote{For example, existing generative AI tools such as GitHub Copilot are in general productivity enhancing because they are designed for specific tasks; Copilot, for example, only completes code and is not an open-ended chatbot.} Second, more detailed data would also shed light on the question of why, after the initial increase in output and quality, we observe a decrease in the effect size for less experienced users in subsequent days. One explanation, in line with the conclusions of \cite{kabir2024} and \cite{li2023}, could be that less experienced users still prefer to use ChatGPT as a support tool because it generates accessible and easy-to-use responses and because the costs of pursuing alternative solutions (e.g., acquiring the necessary coding skills) is relatively high. Finally, our study provides evidence on the productivity effects of (the ban on) generative AI in only the very short run because the ban was short-lived and circumventing it was relatively easy.

\section{Conclusion}
We present novel evidence of the short-term effects of generative AI (ChatGPT) on the productivity of knowledge workers using high-frequency, observational data from over 36,000 software developers in Italy and other European countries. We use the sudden ban on ChatGPT in Italy as a natural experiment and show that the access restriction distorted output quantity and quality. Our results not only present some first empirical evidence of the widespread adoption of ChatGPT in software and code development but also show that the productivity effects of ChatGPT (and restrictions on it) differ by experience level. Our findings have the following policy implications: For some, more complex tasks, generative AI can produce faulty and erroneous output that is difficult to detect, in particular for less experienced individuals. This calls for a more targeted use of the tool in both education and work. AI-based tools that harness the power of LLMs in a more controlled form, that generate a clearly defined output and that are not based on simple text prompts (e.g., GitHub Copilot) offer guard rails to ensure more domain-specific use. Our findings also indicate that even well-intended government-mandated blocking of digital technology (to protect privacy) can lead to short-term output disruptions and costs for society. Sudden bans can be easily circumvented with VPN tools, but these adjustment activities simultaneously distort production processes and negatively impact productivity in professions that rely on the banned technology. Thus, our research also implies that policymakers should consider the potential economic cost of digital technology bans before imposing them.

%We study the consequences of the ban of ChatGPT, a web-based, generative AI technology, in Italy. We compile high-frequency data on GitHub activity from over 36,000 users in Italy and other European countries to measure individual level output of software and code developers. 

%We show that the sudden ban of ChatGPT decreased output by Italian users of around 50\% in the first two days after the ban. We do not find any effects on output after that. This pattern is likely driven by Italian users' efforts to bypass the ban. Using Google search trend and Tor data in a synthetic control design we find that searches for VPN increases by around 52 percentage points in the days after the ban while the usage of Tor bridges increases by 9.4 percentage points. 

%\textbf{[DAVID TO PR: is this still fitting??]}

%Our findings also indicate that government-mandated blocking of digital technology can affect workers differently. While these measures may be well-intended, they are often ineffective and can lead to short-term disruptions in output. Sudden bans can be easily circumvented with VPN tools, but these adjustment activities simultaneously distort production processes and negatively impact productivity in professions that rely on the banned technology. This can ultimately lead to short-run disruptions in economic output. Overall, our research highlights the need for policymakers to consider the potential economic cost of digital technology bans before imposing them.

\end{doublespace}

\clearpage

\bibliographystyle{aer}
\bibliography{bibliography}

\newpage

\begin{table}[H]

\caption{Effect of ChatGPT Ban on GitHub Output \label{tab:did-main}}
\centering
\resizebox{\linewidth}{!}{
\begin{threeparttable}
\begin{tabular}[t]{lllccccc}
\toprule
 &  &  & (1) & (2) & (3) & (4) & (5)\\
\midrule
\addlinespace[0.3em]
\multicolumn{8}{l}{\textbf{A: Output Quantity and Quality}}\\
\hspace{1em} &  & \makecell{ \\ \\ } & \makecell{Output \\ \\ } & \makecell{Output \\ (Shen, 2023) \\} & \makecell{Output \\ (Holub \& \\ Thies, 2023)} & \makecell{Issue \\ closed \\} & \makecell{PR merge \\ ratio \\ }\\
\cmidrule{2-8}
\hspace{1em} & Overall & Treated $\times$ Post & 0.0062 & 0.0102 & 0.0090 & -0.0045** & 0.0045\\

\hspace{1em} & (N = 290,864) &  & (0.0076) & (0.0075) & (0.0080) & (0.0022) & (0.0030)\\
\cmidrule(l{3pt}r{3pt}){3-8}

\hspace{1em} &  & Dep. var. mean & 0.2314 & 0.2283 & 0.2624 & 0.0142 & 0.0359\\
\cmidrule{2-8}
\hspace{1em} & Less experienced & Treated $\times$ Post & 0.0193* & 0.0216** & 0.0203* & -0.0017 & 0.0118***\\

\hspace{1em} & (N = 149,680) &  & (0.0107) & (0.0107) & (0.0111) & (0.0023) & (0.0035)\\
\cmidrule(l{3pt}r{3pt}){3-8}

\hspace{1em} &  & Dep. var. mean & 0.2339 & 0.2316 & 0.2519 & 0.0089 & 0.0258\\
\cmidrule{2-8}
\hspace{1em} & Experienced & Treated $\times$ Post & -0.0079 & -0.0020 & -0.0030 & -0.0077* & -0.0034\\

\hspace{1em} & (N = 141,184) &  & (0.0107) & (0.0106) & (0.0116) & (0.0039) & (0.0052)\\
\cmidrule(l{3pt}r{3pt}){3-8}

\hspace{1em} &  & Dep. var. mean & 0.2287 & 0.2249 & 0.2734 & 0.0198 & 0.0467\\
\cmidrule{2-8}
\hspace{1em} & Pkg. contributor & Treated $\times$ Post & 0.0181 & 0.0164 & 0.0270 & -0.0151* & 0.0113\\

\hspace{1em} & (N = 47,328) &  & (0.0205) & (0.0200) & (0.0215) & (0.0084) & (0.0104)\\
\cmidrule(l{3pt}r{3pt}){3-8}

\hspace{1em} &  & Dep. var. mean & 0.2688 & 0.2644 & 0.3312 & 0.0267 & 0.0615\\
\cmidrule{1-8}
\addlinespace[0.3em]
\multicolumn{8}{l}{\textbf{B: Task Choice and Complexity}}\\
\hspace{1em} &  & \makecell{ \\ \\} & \makecell{PR opened \\ \\ } & \makecell{Avg. lines \\  added per \\ opened PR} & \makecell{Avg. lines \\ added per \\ merged PR} & \makecell{Easy issue \\ closed \\} & \makecell{Interactive \\ Activity \\}\\
\cmidrule{2-8}
\hspace{1em} & Overall & Treated $\times$ Post & 0.0000 & -0.0039 & 0.0220 & -0.0004 & 0.0006\\

\hspace{1em} & (N = 290,864) &  & (0.0034) & (0.0148) & (0.0135) & (0.0004) & (0.0043)\\
\cmidrule(l{3pt}r{3pt}){3-8}

\hspace{1em} &  & Dep. var. mean & 0.0402 & 0.1581 & 0.1439 & 0.0007 & 0.0637\\
\cmidrule{2-8}
\hspace{1em} & Less experienced & Treated $\times$ Post & 0.0040 & 0.0211 & 0.0456*** & -0.0004 & 0.0037\\

\hspace{1em} & (N = 149,680) &  & (0.0040) & (0.0186) & (0.0164) & (0.0004) & (0.0045)\\
\cmidrule(l{3pt}r{3pt}){3-8}

\hspace{1em} &  & Dep. var. mean & 0.0315 & 0.1358 & 0.1148 & 0.0005 & 0.0340\\
\cmidrule{2-8}
\hspace{1em} & Experienced & Treated $\times$ Post & -0.0046 & -0.0324 & -0.0031 & -0.0003 & -0.0028\\

\hspace{1em} & (N = 141,184) &  & (0.0056) & (0.0237) & (0.0221) & (0.0008) & (0.0077)\\
\cmidrule(l{3pt}r{3pt}){3-8}

\hspace{1em} &  & Dep. var. mean & 0.0494 & 0.1817 & 0.1749 & 0.0009 & 0.0953\\
\cmidrule{2-8}
\hspace{1em} & Pkg. contributor & Treated $\times$ Post & -0.0221* & -0.0933* & 0.0226 & -0.0021* & 0.0127\\

\hspace{1em} & (N = 47,328) &  & (0.0117) & (0.0477) & (0.0422) & (0.0011) & (0.0158)\\
\cmidrule(l{3pt}r{3pt}){3-8}

\hspace{1em} &  & Dep. var. mean & 0.0661 & 0.2356 & 0.2210 & 0.0014 & 0.1393\\
\bottomrule
\end{tabular}
\begin{tablenotes}[para]
\item \textit{Notes:} 
\item All specifications include user--fixed effects, day-of-the-week--fixed effects, and a linear time trend for the control and treatment group. The ``Less experienced'' sample includes all Github user accounts created after or in 2017 (median), while the ``Experienced'' sample comprises all GitHub user accounts created before 2017. The ``Pkg. contributor'' sample comprises all GitHub user accounts that are the owner and/or contributor to a (analytical) programming package repository. The number of observations is depicted in parentheses after each sample definition. A log plus one transformation is applied to \textit{Avg. lines added per PR} (opened or merged). Robust standard errors in parentheses are clustered on the user-level: * $p<0.1$, ** $p<0.05$, *** $p<0.01$.
\end{tablenotes}
\end{threeparttable}}
\end{table}

\appendix
\thispagestyle{empty}
% {\noindent \huge \textbf{Supplementary Online Appendix}}

% \parttoc
% \addcontentsline{toc}{section}{Appendix} % Add the appendix text to the document TOC
\part{Supplementary Online Appendix} % Start the appendix part
\parttoc % Insert the appendix TOC

\renewcommand{\thesection}{\Alph{section}}

\numberwithin{table}{section} 
\numberwithin{figure}{section} 
\numberwithin{equation}{section}

% table of contents for Appendix
% \addcontentsline{toc}{section}{Appendix} % Add the appendix text to the document TOC
% \part{} % Start the appendix part
% \parttoc % Insert the appendix TOC

\newpage
\setcounter{page}{1}

\section{Timeline} \label{sec:time-line}

Figure \ref{fig:event-time-line} depicts the timing of the events. The ban was introduced on Friday March 31st 2023. In our baseline analysis, we consider the 4 business days prior to the ban, (MON -- THU, March 27th -- 30th) as the pre-period. To make the analysis, comparable, we ignore the two days following the introduction of the ban which were weekend days (SAT and SUN, April 1st -- 2nd, in gray are \textit{excluded}). The Post period is defined as the first 4 business days, after the introduction of the ban (MON -- THU, April 3rd -- 6th 2023). We further exclude Friday, 7th 2023, because this was Good Friday, a national holiday in all European countries under consideration in our sample.

%%%%%%%%%%%%%%%%%%%%%%%%%%%%%%%%%%%%%%%%%%%%%%%%%%%%%%%%%%%%%%%%%%%%%
% Timeline Graph
%%%%%%%%%%%%%%%%%%%%%%%%%%%%%%%%%%%%%%%%%%%%%%%%%%%%%%%%%%%%%%%%%%%%%

\begin{figure}[!h]
    \centering
   \caption{Time Line}
    \label{fig:event-time-line}
    \begin{tikzpicture}[scale = 0.9]
       % draw horizontal line
        \draw (0,0) -- (1,0);
        \draw (0,0) -- (2.5,0) node[pos=.5,above] {PRE Period};
        \draw (2.5,0) -- (5,0);
        \draw (5,0) -- (7.5,0) node[pos=.5,above] {\textit{Ban Intro/WE (Exc.)}};
         \draw (7.5,0) -- (10,0);
        \draw (10,0) -- (13,0)  node[pos=.5,above] {POST Period};
                  \draw (13,0) -- (16,0)  node[pos=.5,above] {\textit{Easter (Exc.)}};

        %\draw (9,0) -- (14,0) node[pos=.5,above] {(post-event window]};
        %\draw (11,0) -- (13,0);
 % draw rectangle for Ban Intro/WE
        \fill[gray] (5, -0.1) rectangle (7.5, 0.1);
        \fill[gray] (13, -0.1) rectangle (16, 0.1);
        % draw vertical line
        \foreach \x in {0,2.5,5,7.5,10,13}
                \draw (\x cm,3pt) -- (\x cm,-3pt);
            
        % draw nodes
        \draw (0,0) node[below=3pt] {MON 27/03} node[above=3pt] {$ $};
       % \draw (2.5,0) node[below=3pt] {TUE 28/03} node[above=3pt] {$ $};
        \draw (2.5,0) node[below=3pt] {THU 30/03} node[above=3pt] {$ $};
        \draw (5,0) node[below=3pt] {FRI 31/03} node[above=3pt] {$ $};
        \draw (7.5,0) node[below=3pt] {SUN 02/04} node[above=3pt] {$ $};
        \draw (10,0) node[below=3pt] {MON 03/04} node[above=3pt] {$ $};
        % \draw (12.5,0) node[below=3pt] {TUE 04/04} node[above=3pt] {$ $};
        \draw (13,0) node[below=3pt] {THU 06/04} node[above=3pt] {$ $};

    \end{tikzpicture}
\end{figure}
 
\clearpage

\section{Data Appendix}

\subsection{Github User Location} \label{sec:github-user}

\begin{wrapfigure}{L}{0.5\textwidth}
  \caption{GitHub User Profile}
  \begin{center}
    \includegraphics[width=0.48\textwidth]{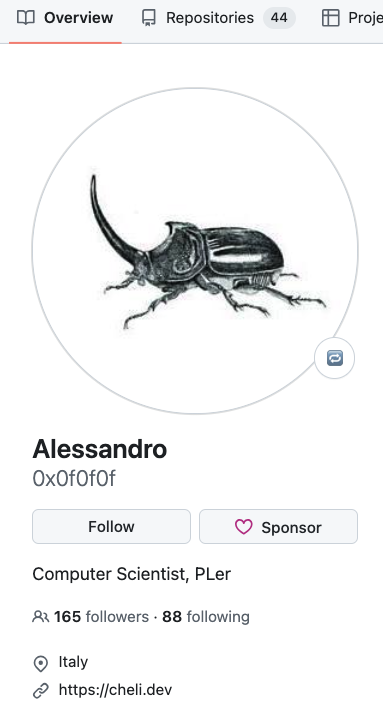}
  \label{fig:distribution-counts}    \captionsetup{justification=justified, singlelinecheck=off} 
\caption*{\footnotesize \textit{Notes:} Profile last accessed on 07 May 2024 at \url{https://github.com/0x0f0f0f}.}
  \end{center}
\end{wrapfigure}

We scrape, among others, the following key attributes at the user-level for a defined set of locations (\texttt{location}) and span of time-period (i.e. \texttt{createdAt} from 1 Jan 2009 until 11 Apr 2023) from the \href{https://docs.github.com/en/graphql}{GitHub GraphQL API}:\footnote{For more details on the user-level attributes please refer to the official GitHub documentation: \url{https://docs.github.com/en/graphql/reference/objects\#user}.}
\begin{itemize}
    \item \texttt{login}: The user's username (e.g. \texttt{0x0f0f0f}).
    % \item \texttt{name}: The user's full name.
    \item \texttt{location}: The user's location (e.g. \texttt{Italy}).
    % \item  \texttt{company}: The user's company name (e.g. \texttt{Harvard University}).
    \item \texttt{followers}: The number of followers the user has (e.g. \texttt{165}).
    \item \texttt{following}: The numbers of users the user is following (e.g. \texttt{88}).
    \item \texttt{repositories}: A list of user's (public) repositories (e.g. \texttt{44}).
    \item \texttt{createdAt}: The date and time the user's account was created.
    % \item \texttt{updatedAt}: The date and time the user's account was last updated.
\end{itemize}

For more details on the scraping procedure (e.g. the set of locations) and the \texttt{python} implementation please refer to the public GitHub repository  \url{https://GitHub.com/sodalabsio/GitHub_scrape}.

\clearpage

\subsection{Definition and Construction of Outcome Variables} \label{sec:var-defs}

\begin{ThreePartTable}

% \begin{TableNotes}
% \item \textit{Notes:} The first column presents the variable name, and the second column provides a detailed description of how each variable is defined. The SQL Google BigQuery code to retrieve the required data is presented in brackets. The keywords to define an issue as ``easy'' are \texttt{good first issues, good first bug, good-first, documentation, polish, cleanup, simple, easy, small, trivial, minor help, wanted, junior job, newcomer, starter, beginner, newbie, novice, low hanging, low-hanging} \citepappx[cf.][]{holub2023}. ``Productive'' user activity comprises the following events: \texttt{PullRequestEvent, PullRequestReviewEvent, PullRequestReviewCommentEvent,PushEvent, ReleaseEvent, CreateEvent, IssueEvent}. 
% \end{TableNotes}
\begin{longtable}[t]{>{\raggedright\arraybackslash}p{4.5cm}>{\raggedright\arraybackslash}p{10.5cm}}
\caption{Variable Definitions} \label{tab:var-defs}\\
\toprule
Variable & Description \\
\midrule
\endfirsthead
\caption[]{Variable Definitions \textit{(continued)}}\\
\toprule
Variable & Description \\
\midrule
\endhead

\endfoot
\bottomrule
%\insertTableNotes
\endlastfoot
% Output Quantity ---------------------------------------------------
\textbf{A -- Output Quantity} & \\
\hline
% Events
\cellcolor{gray!6}{Events}  & \cellcolor{gray!6}{Sum of all 15 GitHub event types, i.e., \texttt{CommitCommentEvent}, \texttt{CreateEvent}, \texttt{DeleteEvent}, \texttt{ForkEvent}, \texttt{GollumEvent}, \texttt{IssuesEvent}, \texttt{IssueCommentEvent}, \texttt{MemberEvent}, \texttt{PublicEvent}, \texttt{PullRequestEvent}, \texttt{PullRequestReviewEvent}, \texttt{PullRequestReviewCommentEvent}, \texttt{PushEvent}, \texttt{ReleaseEvent}, \texttt{WatchEvent}}\\
% Output 
Output & \# Commits [\texttt{PushEvent\$.size}] + \# Issues closed [\texttt{IssuesEvent\$.action == 'closed'}]+ \# Pull requests closed [\texttt{PullRequestEvent\$.action == 'closed'}] + \# Releases [\texttt{ReleaseEvent}]\\
% Output (Shen, 2023)
\cellcolor{gray!6}{Outputs \citepappx{shen2023}}  & \cellcolor{gray!6}{\# Commits [\texttt{PushEvent\$.size}] + \# Pull requests [\texttt{PullRequestEvent}]}\\
% Output (Holub and Thiel, 2023)
Outputs \citepappx{holub2023} & \# Commits [\texttt{PushEvent\$.size}] + \# Comments on issues [\texttt{IssueCommentEvent}]+ \# Comments on pull requests [\texttt{PullRequestReviewCommentEvent}] + \# Comments on commits [\texttt{CommitCommentEvent}] + \# Pull requests [\texttt{PullRequestEvent}] + \# Issues [\texttt{IssuesEvent}]\\
% Issues closed
\cellcolor{gray!6}{Issues closed}  & \cellcolor{gray!6}{\# Issues closed [\texttt{IssuesEvent\$.action == 'closed'}]} \\
% Commit
Commits & \# Commits [\texttt{PushEvent\$.size}]\\
% Releases
% \cellcolor{gray!6}{Releases} & \cellcolor{gray!6}{\# Releases [\texttt{ReleaseEvent}]}\\
% Output Quality -----------------------------------------------------
\hline
\textbf{B -- Output Quality} & \\
\hline
% Pull Requests
% Pull Requests (PRs) merged & \# Pull requests merged [\texttt{PullRequest\$.pull\_request.merged == 'true' \& PullRequestEvent\$.action == 'closed'}]\\
% \cellcolor{gray!6}{Bug fix ratio}  & \cellcolor{gray!6}{(\# Issues closed with ``error'' or ``bug'' label [\texttt{IssuesEvent\$.action == 'closed' \& REGEXP\_CONTAINS(IssuesEvent\$.label, 'error|bug'\}]) / (\# Issues closed [\texttt{IssuesEvent\$.action == 'closed'}])}}\\
Pull Requests (PRs) merged & \# Closed pull requests that were merged [\texttt{PullRequestEvent\$.pull\_request.merged == 'true' \& PullRequestEvent\$.action == 'closed'}] \\
\cellcolor{gray!6}{PR merge ratio}  & \cellcolor{gray!6}{( \# Closed pull requests that were merged [\texttt{PullRequestEvent\$.pull\_request.merged == 'true' \& PullRequestEvent\$.action == 'closed'}]) / (\# Pull Requests closed [\texttt{PullRequestEvent\$.action == 'closed'}])}\\
\cellcolor{gray!6}{PR merge ratio \citep{holub2023}}  & \cellcolor{gray!6}{( \# Opened pull requests that were merged  [\texttt{PullRequest\$.pull\_request.merged == 'true' \& PullRequestEvent\$.action == 'closed' \& PullRequestEvent\$.action == 'opened'}]) / (\# Pull Requests opened [\texttt{PullRequestEvent\$.action == 'opened'}])}\\
\hline
\multicolumn{2}{l}{\textbf{C -- Task Choice and Complexity}}\\
\hline
% Task Complexity - Files edited
Pull Requests (PRs) opened & \# Pull Requests opened [\texttt{PullRequestEvent\$.action == 'opened'}] \\
\cellcolor{gray!6}{Avg. files edited per merged PR}  & \cellcolor{gray!6}{ Average \# files edited per merged pull request [\texttt{AVG(PullRequestEvent\$.pull\_request.changed\_files) IF PullRequestEvent\$.pull\_request.merged == 'true' \& PullRequestEvent\$.action == 'closed'}]} \\
% Task Complexity - Lines added
Avg. Lines added per merged PR  & Average \# lines added per merged pull request [\texttt{AVG(PullRequestEvent\$.pull\_request.additions) IF PullRequestEvent\$.pull\_request.merged == 'true' \& PullRequestEvent\$.action == 'closed'}] \\
% Task Complexity - Lines added
\cellcolor{gray!6}{Avg. Lines added per opened PR}  & \cellcolor{gray!6}{Average \# lines added per opened pull request [\texttt{AVG(PullRequestEvent\$.pull\_request.additions) IF PullRequestEvent\$.action == 'opened'}]} \\
%  Easy issue closed
Easy issue closed & \# Easy issues closed [\texttt{IssuesEvent\$.action == 'closed'}] \\
% Interactive Activity
\cellcolor{gray!6}{Interactive Activity}  & \cellcolor{gray!6}{\# Comments on issues [\texttt{IssueCommentEvent}]+ \# Comments on pull requests [\texttt{PullRequestReviewCommentEvent}] + \# Comments on commits [\texttt{CommitCommentEvent}]} \\
\hline
\textbf{D -- User Activity} & \\
\hline
``Work'' hours & Time difference (in hours) between first and last activity (any or ``prodctive'')]\\
\cellcolor{gray!6}{First activity}  & \cellcolor{gray!6}{Clock Hour of first activity (any or ``prodctive'')} \\
Last activity  & Clock Hour of last activity (any or ``prodctive'')\\
\cellcolor{gray!6}{Unusual activity (10th \%tile)}  & \cellcolor{gray!6}{Dummy = 1 if activity (any or ``prodctive'') before 6am (10th percentile of \emph{First activity}) or after 9pm (90th percentile of  \emph{Last activity})}\\
Unusual activity (25th \%tile) & Dummy = 1 if activity (any or ``prodctive'') before 8am (25th percentile of \emph{First activity}) or after 7pm (75th percentile of \emph{Last activity})\\
% \textbf{D -- Placebo} & \\
% \hline
% \cellcolor{gray!6}{Wiki pages}  & \cellcolor{gray!6}{\# Wiki pages created  [\texttt{GollumEvent}]} \\
% Public repos & \# Repositories made public [\texttt{PublicEvent}] \\
\end{longtable}
\end{ThreePartTable}
\vspace*{-\baselineskip}
\noindent{\justifying \footnotesize \textit{Notes:} \hspace{0.5em} The first column presents the variable name, and the second column provides a detailed description of how each variable is defined. The SQL Google BigQuery code to retrieve the required data is presented in brackets. The keywords to define an issue as ``easy'' are \texttt{good first issues}, \texttt{good first bug}, \texttt{good-first}, \texttt{documentation}, \texttt{ polish}, \texttt{cleanup}, \texttt{simple}, \texttt{easy}, \texttt{small}, \texttt{trivial}, \texttt{minor help wanted}, \texttt{junior job}, \texttt{newcomer}, \texttt{starter}, \texttt{beginner}, \texttt{newbie}, \texttt{novice}, \texttt{low hanging}, \texttt{low-hanging} \citepappx[cf.][]{holub2023}. ``Productive'' user activity comprises the following events: \texttt{PullRequestEvent}, 
 \texttt{PullRequestReviewEvent}, \texttt{PullRequestReviewCommentEvent}, \texttt{PushEvent}, \texttt{ReleaseEvent}, \texttt{CreateEvent}, \texttt{IssueEvent}.}

\clearpage

\section{GitHub User-Level Data}

\subsection{Descriptive Statistics}

\begin{table}[H]

\caption{Descriptive Statistics \label{tab:summary-user}}
\centering
\resizebox{\linewidth}{!}{
\begin{threeparttable}
\begin{tabular}{lrrrrrrrr}
\toprule
\multicolumn{1}{c}{ } & \multicolumn{2}{c}{Overall} & \multicolumn{2}{c}{Less Experienced} & \multicolumn{2}{c}{Experienced} & \multicolumn{2}{c}{Pkg. contributor} \\
\multicolumn{1}{c}{ } & \multicolumn{2}{c}{(N = 36,358)} & \multicolumn{2}{c}{(N = 18,710)} & \multicolumn{2}{c}{(N = 17,648)} & \multicolumn{2}{c}{(N = 5,916)} \\
\cmidrule(l{3pt}r{3pt}){2-3} \cmidrule(l{3pt}r{3pt}){4-5} \cmidrule(l{3pt}r{3pt}){6-7} \cmidrule(l{3pt}r{3pt}){8-9}
  & Mean & SD & Mean & SD & Mean & SD & Mean & SD\\
\midrule
\addlinespace[0.3em]
\multicolumn{9}{l}{\textbf{A - User-Day--Level (N = 290,864)}}\\
\hline
\hspace{1em}Output & 0.2314 & 0.4217 & 0.2339 & 0.4233 & 0.2287 & 0.4200 & 0.2688 & 0.4433\\
\hspace{1em}Output (Shen, 2023) & 0.2283 & 0.4198 & 0.2316 & 0.4218 & 0.2249 & 0.4175 & 0.2644 & 0.4410\\
\hspace{1em}Output (Holub and Thies, 2023) & 0.2624 & 0.4399 & 0.2519 & 0.4341 & 0.2734 & 0.4457 & 0.3312 & 0.4706\\
\hspace{1em}Issue closed & 0.0142 & 0.1183 & 0.0089 & 0.0938 & 0.0198 & 0.1394 & 0.0267 & 0.1611\\
\hspace{1em}PR merge ratio & 0.0359 & 0.1846 & 0.0258 & 0.1572 & 0.0467 & 0.2092 & 0.0615 & 0.2381\\
\hspace{1em}PR opened & 0.0402 & 0.1964 & 0.0315 & 0.1746 & 0.0494 & 0.2168 & 0.0661 & 0.2484\\
\hspace{1em}Avg. lines added per opened PR & 0.1581 & 0.9003 & 0.1358 & 0.8648 & 0.1817 & 0.9359 & 0.2356 & 1.0474\\
\hspace{1em}Avg. lines added per merged PR & 0.1439 & 0.8563 & 0.1148 & 0.7928 & 0.1749 & 0.9179 & 0.2210 & 1.0124\\
\hspace{1em}Easy issue closed & 0.0007 & 0.0263 & 0.0005 & 0.0219 & 0.0009 & 0.0302 & 0.0014 & 0.0367\\
\hspace{1em}Interactive Activity & 0.0637 & 0.2442 & 0.0340 & 0.1811 & 0.0953 & 0.2936 & 0.1393 & 0.3463\\
\addlinespace[0.3em]
\hline
\multicolumn{9}{l}{\textbf{B - User--Level (N = 36,358)}}\\
\hline
\hspace{1em} & Mean & SD & Min & Median & Max &  &  & \\
\cmidrule(l{3pt}r{3pt}){1-6}
\hspace{1em}User creation year & 2016.55 & 3.77 & 2009 & 2017 & 2023 &  &  & \\
\hspace{1em}Experienced & 0.49 & 0.50 & 0 & 0 & 1 &  &  & \\
\hspace{1em}Pkg. contributions & 49.81 & 447.12 & 0 & 0 & 19638 &  &  & \\
\hspace{1em}Pkg. owner & 0.05 & 0.22 & 0 & 0 & 1 &  &  & \\
\hspace{1em}Followers & 29.49 & 203.22 & 0 & 6 & 17421 &  &  & \\
\hspace{1em}Following & 19.67 & 185.41 & 0 & 5 & 28300 &  &  & \\
\hspace{1em}Repositories & 29.90 & 54.53 & 0 & 17 & 3900 &  &  & \\
\hspace{1em}Total events & 11.93 & 18.87 & 1 & 5 & 140 &  &  & \\
\bottomrule
\end{tabular}
\begin{tablenotes}[para]
\item \textit{Notes:} 
\item Panel A presents descriptive statistics for the baseline sample period Pre 27-30.03 -- Post 03-06.04. The ``Less experienced'' sample includes all GitHub user accounts created after or in 2017 (median), while the ``Experienced'' sample comprises all GitHub user accounts created before 2017. The ``Pkg. contributor'' sample comprises all GitHub user accounts that are the owner and/or contributor to a (analytical) programming package repository. The number of unique GitHub user accounts for the entire baseline sample (``Overall'') and each of the subsamples is presented in parentheses below. A log plus one transformation is applied to \textit{Avg. lines added per PR} (opened or merged). Panel B provides information on the individual characteristics of all GitHub user accounts in the baseline sample.
\end{tablenotes}
\end{threeparttable}}
\end{table}

\begin{figure}[h!]
    \centering
    \caption{Distribution of Output Quantities}
    \includegraphics[width=\textwidth]{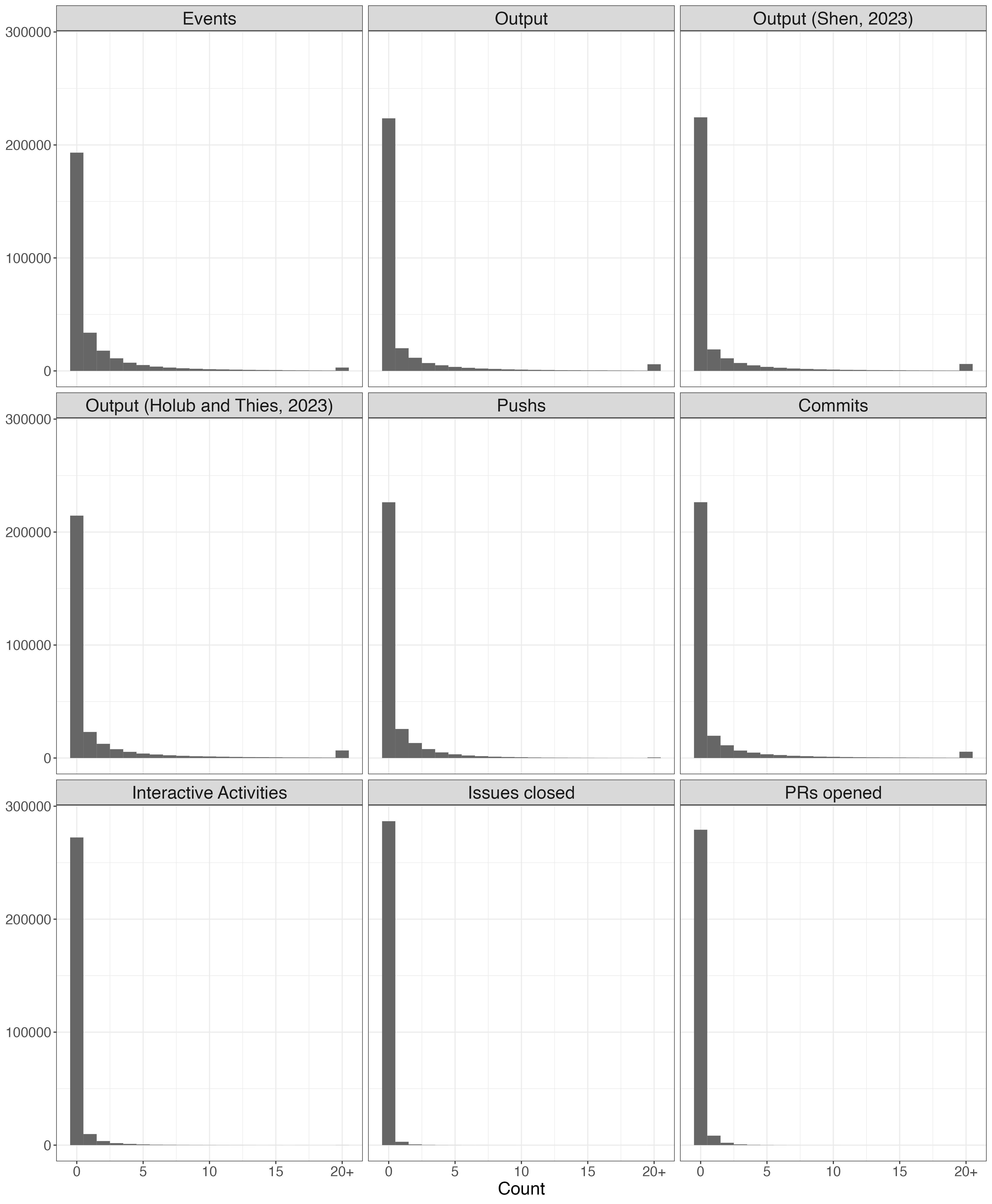}
    \label{fig:distribution-counts}
    \captionsetup{justification=justified, singlelinecheck=off} 
\caption*{\footnotesize \textit{Notes:} Daily counts of each action type at the user level for the sample period of March 27--30 (Pre) -- April 3--6 (Post) are presented. Counts above 20 are binned and labelled $20+$.}
\end{figure}

\clearpage

\begin{figure}[h!]
    \centering
    \caption{Distribution of Activity across Hours of the Day in Italy Pre-/Post-ChatGPT--Ban}
    \includegraphics[width=\textwidth]{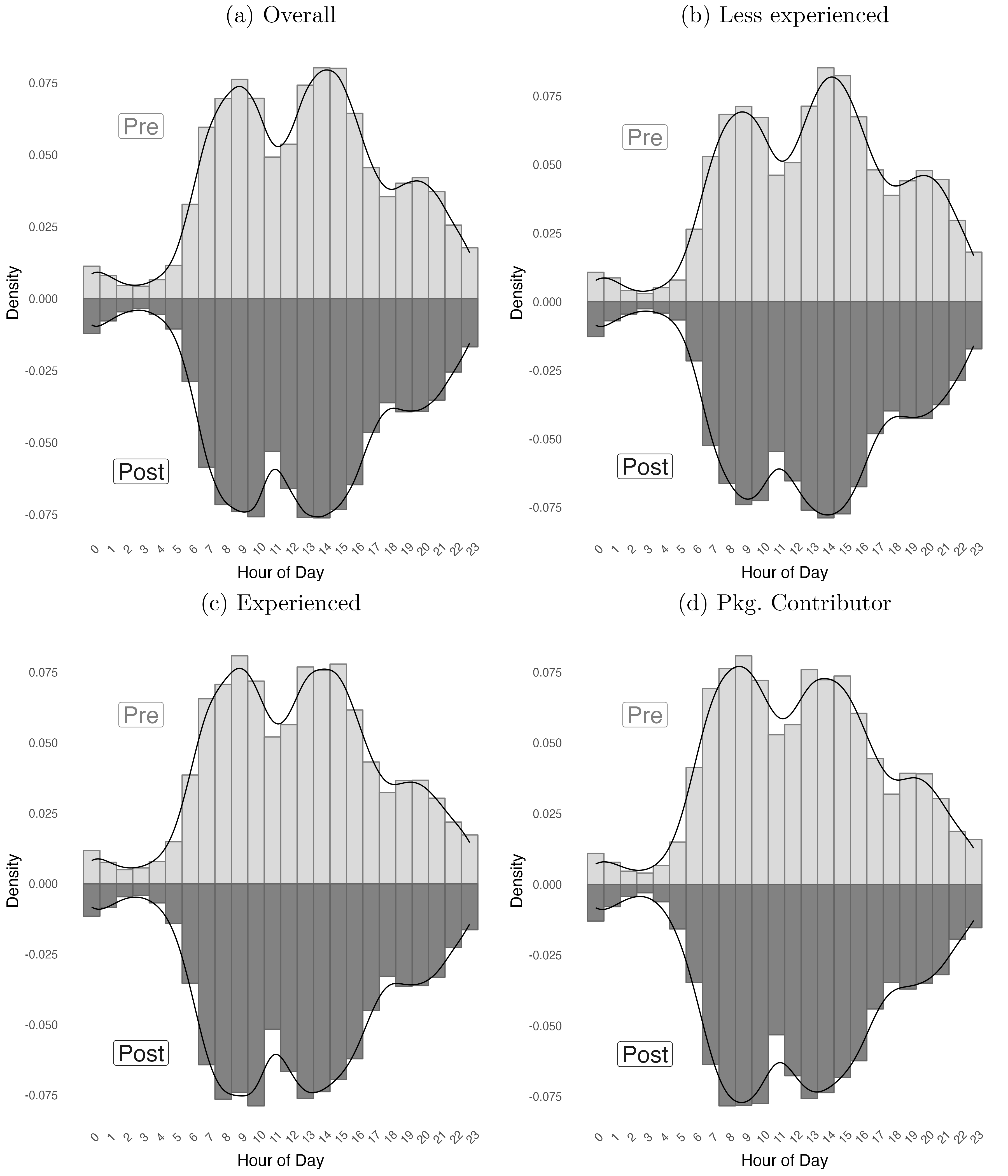}
    \label{fig:distribution-work-activity-Italy}
    \captionsetup{justification=justified, singlelinecheck=off} 
\caption*{\footnotesize \textit{Notes:} The distribution of (any) activity by GitHub users across hours of the day is presented separately for the four days pre- and post-ban.}
\end{figure}

\begin{figure}[h!]
    \centering
    \caption{Distribution of Activity across Hours of the Day in Control Group Pre-/Post-ChatGPT--Ban}
    \includegraphics[width=\textwidth]{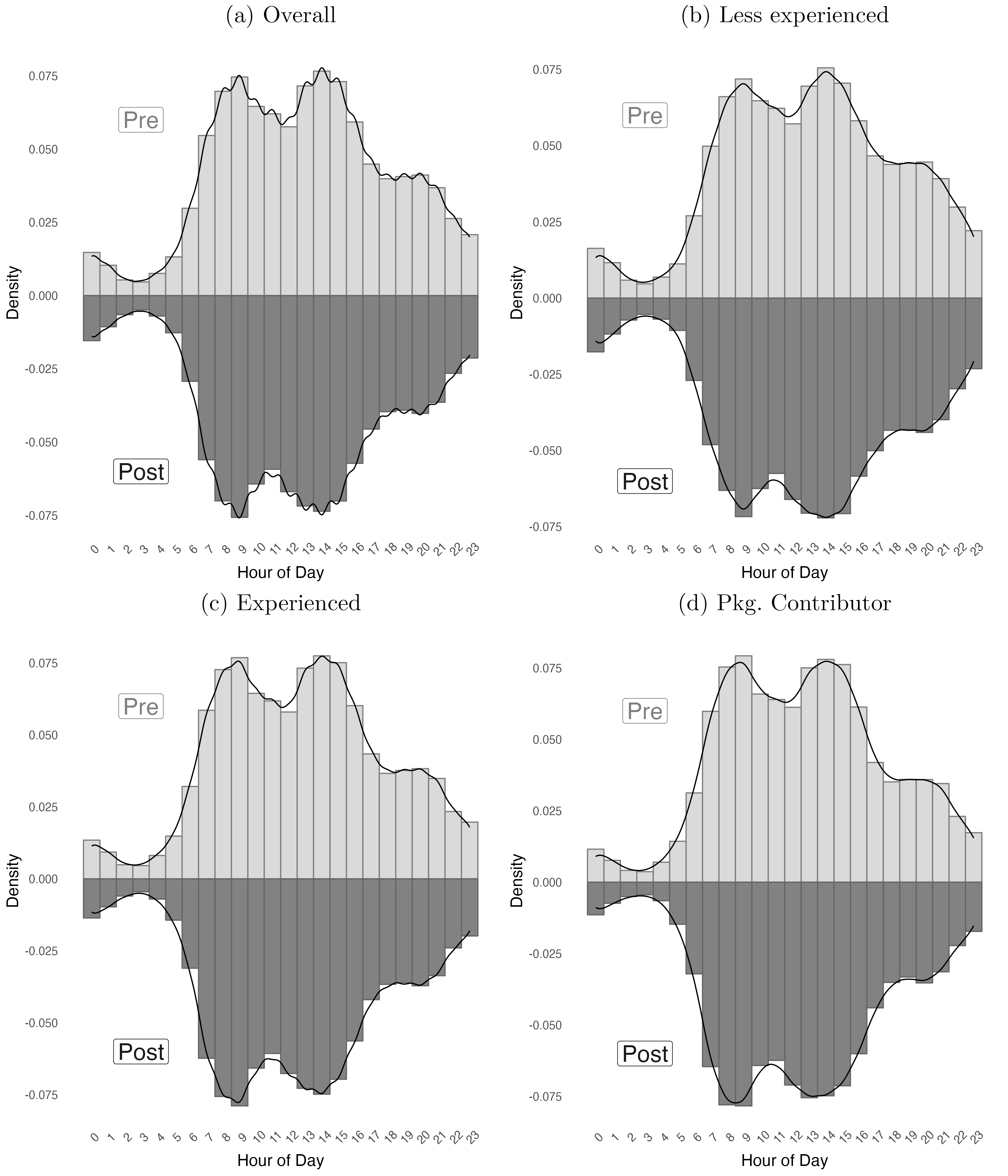}
    \label{fig:distribution-work-activity-Italy}
    \captionsetup{justification=justified, singlelinecheck=off} 
\caption*{\footnotesize \textit{Notes:} The distribution of (any) activity by GitHub users across hours of the day is presented separately for the four days pre- and post-ban.}
\end{figure}

\subsection{Additional Results} \label{sec:user-add-results}

\begin{table}[H]

\caption{DID Specification without Linear Time-Trends \label{tab:did-wo-trend}}
\centering
\resizebox{\linewidth}{!}{
\begin{threeparttable}
\begin{tabular}[t]{lllccccc}
\toprule
 &  &  & (1) & (2) & (3) & (4) & (5)\\
\midrule
\addlinespace[0.3em]
\multicolumn{8}{l}{\textbf{A: Output Quantity and Quality}}\\
\hspace{1em} &  & \makecell{ \\ \\ } & \makecell{Output \\ \\ } & \makecell{Output \\ (Shen, 2023) \\} & \makecell{Output \\ (Holub \& \\ Thies, 2023} & \makecell{Issue \\ closed \\} & \makecell{PR merge \\ ratio \\ }\\
\cmidrule{2-8}
\hspace{1em} & Overall & Treated $\times$ Post & 0.0116*** & 0.0128*** & 0.0147*** & -0.0005 & 0.0056***\\

\hspace{1em} &  (N = 290,864)  &  & (0.0043) & (0.0042) & (0.0045) & (0.0011) & (0.0015)\\
\cmidrule(l{3pt}r{3pt}){3-8}

\hspace{1em} &  & Dep. var. mean & 0.2314 & 0.2283 & 0.2624 & 0.0142 & 0.0359\\
\cmidrule{2-8}
\hspace{1em} & Less experienced  & Treated $\times$ Post & 0.0177*** & 0.0177*** & 0.0186*** & 0.0027** & 0.0095***\\

\hspace{1em} & (N = 149,680) &  & (0.0061) & (0.0061) & (0.0062) & (0.0013) & (0.0018)\\
\cmidrule(l{3pt}r{3pt}){3-8}

\hspace{1em} &  & Dep. var. mean & 0.2339 & 0.2316 & 0.2519 & 0.0089 & 0.0258\\
\cmidrule{2-8}
\hspace{1em} & Experienced & Treated $\times$ Post & 0.0052 & 0.0079 & 0.0108* & -0.0043** & 0.0013\\

\hspace{1em} &  (N = 141,184) &  & (0.0059) & (0.0059) & (0.0064) & (0.0020) & (0.0025)\\
\cmidrule(l{3pt}r{3pt}){3-8}

\hspace{1em} &  & Dep. var. mean & 0.2287 & 0.2249 & 0.2734 & 0.0198 & 0.0467\\
\cmidrule{2-8}
\hspace{1em} & Pkg. contributor  & Treated $\times$ Post & 0.0148 & 0.0150 & 0.0194* & -0.0095** & -0.0001\\

\hspace{1em} & (N = 47,328) &  & (0.0109) & (0.0108) & (0.0115) & (0.0040) & (0.0049)\\
\cmidrule(l{3pt}r{3pt}){3-8}

\hspace{1em} &  & Dep. var. mean & 0.2688 & 0.2644 & 0.3312 & 0.0267 & 0.0615\\
\cmidrule{1-8}
\addlinespace[0.3em]
\multicolumn{8}{l}{\textbf{B: Task Choice and Complexity}}\\
\hspace{1em} &  & \makecell{ \\ \\} & \makecell{PR opened \\ \\ } & \makecell{Avg. lines \\  added per \\ opened PR} & \makecell{Avg. lines \\ added per \\ merged PR} & \makecell{Easy issue \\ closed \\} & \makecell{Interactive \\ Activity \\}\\
\cmidrule{2-8}
\hspace{1em} & Overall & Treated $\times$ Post & 0.0035** & 0.0180** & 0.0249*** & 0.0002 & 0.0011\\

\hspace{1em} &  &  & (0.0017) & (0.0078) & (0.0070) & (0.0002) & (0.0022)\\
\cmidrule(l{3pt}r{3pt}){3-8}

\hspace{1em} & (N = 290,864) & Dep. var. mean & 0.0402 & 0.1581 & 0.1439 & 0.0007 & 0.0637\\
\cmidrule{2-8}
\hspace{1em} & Less experienced  & Treated $\times$ Post & 0.0056*** & 0.0322*** & 0.0402*** & 0.0002 & 0.0016\\

\hspace{1em} & (N = 149,680) &  & (0.0021) & (0.0100) & (0.0087) & (0.0002) & (0.0024)\\
\cmidrule(l{3pt}r{3pt}){3-8}

\hspace{1em} &  & Dep. var. mean & 0.0315 & 0.1358 & 0.1148 & 0.0005 & 0.0340\\
\cmidrule{2-8}
\hspace{1em} & Experienced & Treated $\times$ Post & 0.0012 & 0.0025 & 0.0079 & 0.0001 & 0.0004\\

\hspace{1em} & (N = 141,184) &  & (0.0029) & (0.0121) & (0.0112) & (0.0004) & (0.0039)\\
\cmidrule(l{3pt}r{3pt}){3-8}

\hspace{1em} &  & Dep. var. mean & 0.0494 & 0.1817 & 0.1749 & 0.0009 & 0.0953\\
\cmidrule{2-8}
\hspace{1em} & Pkg. contributor & Treated $\times$ Post & -0.0083 & -0.0269 & -0.0006 & -0.0010* & 0.0005\\

\hspace{1em} & (N = 47,328) &  & (0.0057) & (0.0240) & (0.0219) & (0.0006) & (0.0081)\\
\cmidrule(l{3pt}r{3pt}){3-8}

\hspace{1em} &  & Dep. var. mean & 0.0661 & 0.2356 & 0.2210 & 0.0014 & 0.1393\\
\bottomrule
\end{tabular}
\begin{tablenotes}[para]
\item \textit{Notes:} 
\item All specifications include user--fixed effects and day-of-the-week--fixed effects. The ``Less experienced'' sample includes all Github user accounts created after or in 2017 (median), while the ``Experienced'' sample comprises all GitHub user accounts created before 2017. The ``Pkg. contributor'' sample comprises all GitHub user accounts that are the owner and/or contributor to a (analytical) programming package repository. The number of observations is depicted in parentheses after each sample definition. A log plus one transformation is applied to \textit{Avg. lines added per PR} (opened or merged). Robust standard errors in parentheses are clustered on the user-level: * $p<0.1$, ** $p<0.05$, *** $p<0.01$.
\end{tablenotes}
\end{threeparttable}}
\end{table}

\begin{table}[H]

\caption{DID Specification with Country-Specific Linear Time-Trends \label{tab:did-ctry-trend}}
\centering
\resizebox{\linewidth}{!}{
\begin{threeparttable}
\begin{tabular}[t]{lllccccc}
\toprule
 &  &  & (1) & (2) & (3) & (4) & (5)\\
\midrule
\addlinespace[0.3em]
\multicolumn{8}{l}{\textbf{A: Output Quantity and Quality}}\\
\hspace{1em} &  & \makecell{ \\ \\ } & \makecell{Output \\ \\ } & \makecell{Output \\ (Shen, 2023) \\} & \makecell{Output \\ (Holub \& \\ Thies, 2023} & \makecell{Issue \\ closed \\} & \makecell{PR merge \\ ratio \\ }\\
\cmidrule{2-8}
\hspace{1em} & Overall & Treated $\times$ Post & 0.0062 & 0.0102 & 0.0090 & -0.0045** & 0.0045\\

\hspace{1em} & (N = 290,864) &  & (0.0076) & (0.0075) & (0.0080) & (0.0022) & (0.0030)\\
\cmidrule(l{3pt}r{3pt}){3-8}

\hspace{1em} &  & Dep. var. mean & 0.2314 & 0.2283 & 0.2624 & 0.0142 & 0.0359\\
\cmidrule{2-8}
\hspace{1em} & Less experienced & Treated $\times$ Post & 0.0193* & 0.0216** & 0.0203* & -0.0017 & 0.0118***\\

\hspace{1em} & (N = 149,680) &  & (0.0107) & (0.0107) & (0.0111) & (0.0023) & (0.0035)\\
\cmidrule(l{3pt}r{3pt}){3-8}

\hspace{1em} &  & Dep. var. mean & 0.2339 & 0.2316 & 0.2519 & 0.0089 & 0.0258\\
\cmidrule{2-8}
\hspace{1em} & Experienced & Treated $\times$ Post & -0.0079 & -0.0020 & -0.0030 & -0.0077* & -0.0034\\

\hspace{1em} & (N = 141,184) &  & (0.0107) & (0.0106) & (0.0116) & (0.0039) & (0.0052)\\
\cmidrule(l{3pt}r{3pt}){3-8}

\hspace{1em} &  & Dep. var. mean & 0.2287 & 0.2249 & 0.2734 & 0.0198 & 0.0467\\
\cmidrule{2-8}
\hspace{1em} & Pkg. contributor & Treated $\times$ Post & 0.0181 & 0.0164 & 0.0270 & -0.0151* & 0.0113\\

\hspace{1em} & (N = 47,328) &  & (0.0205) & (0.0200) & (0.0215) & (0.0084) & (0.0104)\\
\cmidrule(l{3pt}r{3pt}){3-8}

\hspace{1em} &  & Dep. var. mean & 0.2688 & 0.2644 & 0.3312 & 0.0267 & 0.0615\\
\cmidrule{1-8}
\addlinespace[0.3em]
\multicolumn{8}{l}{\textbf{B: Task Choice and Complexity}}\\
\hspace{1em} &  & \makecell{ \\ \\} & \makecell{PR opened \\ \\ } & \makecell{Avg. lines \\  added per \\ opened PR} & \makecell{Avg. lines \\ added per \\ merged PR} & \makecell{Easy issue \\ closed \\} & \makecell{Interactive \\ Activity \\}\\
\cmidrule{2-8}
\hspace{1em} & Overall & Treated $\times$ Post & 0.0000 & -0.0039 & 0.0220 & -0.0004 & 0.0006\\

\hspace{1em} & (N = 290,864) &  & (0.0034) & (0.0148) & (0.0135) & (0.0004) & (0.0043)\\
\cmidrule(l{3pt}r{3pt}){3-8}

\hspace{1em} &  & Dep. var. mean & 0.0402 & 0.1581 & 0.1439 & 0.0007 & 0.0637\\
\cmidrule{2-8}
\hspace{1em} & Less experienced & Treated $\times$ Post & 0.0040 & 0.0211 & 0.0456*** & -0.0004 & 0.0037\\

\hspace{1em} & (N = 149,680) &  & (0.0040) & (0.0186) & (0.0164) & (0.0004) & (0.0045)\\
\cmidrule(l{3pt}r{3pt}){3-8}

\hspace{1em} &  & Dep. var. mean & 0.0315 & 0.1358 & 0.1148 & 0.0005 & 0.0340\\
\cmidrule{2-8}
\hspace{1em} & Experienced & Treated $\times$ Post & -0.0046 & -0.0324 & -0.0031 & -0.0003 & -0.0028\\

\hspace{1em} &  (N = 141,184) &  & (0.0056) & (0.0237) & (0.0221) & (0.0008) & (0.0077)\\
\cmidrule(l{3pt}r{3pt}){3-8}

\hspace{1em} &  & Dep. var. mean & 0.0494 & 0.1817 & 0.1749 & 0.0009 & 0.0953\\
\cmidrule{2-8}
\hspace{1em} & Pkg. contributor & Treated $\times$ Post & -0.0221* & -0.0933* & 0.0226 & -0.0021* & 0.0127\\

\hspace{1em} &  (N = 47,328) &  & (0.0117) & (0.0477) & (0.0422) & (0.0011) & (0.0158)\\
\cmidrule(l{3pt}r{3pt}){3-8}

\hspace{1em} &  & Dep. var. mean & 0.0661 & 0.2356 & 0.2210 & 0.0014 & 0.1393\\
\bottomrule
\end{tabular}
\begin{tablenotes}[para]
\item \textit{Notes:} 
\item All specifications include user--fixed effects, day-of-the-week--fixed effects and country-specific linear time trends. The ``Less experienced'' sample includes all Github user accounts created after or in 2017 (median), while the ``Experienced'' sample comprises all GitHub user accounts created before 2017. The ``Pkg. contributor'' sample comprises all GitHub user accounts that are the owner and/or contributor to a (analytical) programming package repository. The number of observations is depicted in parentheses after each sample definition. A log plus one transformation is applied to \textit{Avg. lines added per PR} (opened or merged). Robust standard errors in parentheses are clustered on the user-level: * $p<0.1$, ** $p<0.05$, *** $p<0.01$.
\end{tablenotes}
\end{threeparttable}}
\end{table}

\begin{landscape}
\begin{figure}[H]
\centering 
\caption{Experienced Users -- Event-Study Estimates}
\label{fig:event-study-exp}
\includegraphics[width=1.5\textwidth]{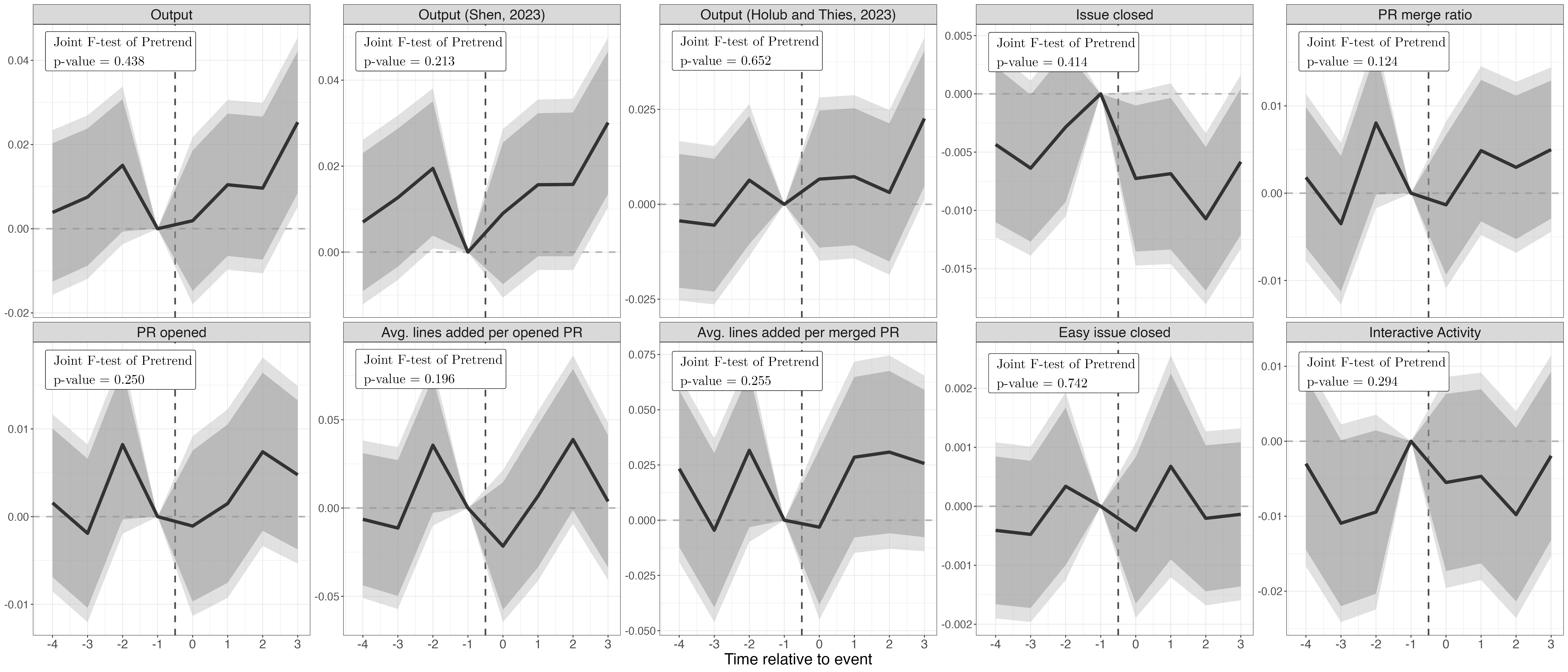}
\captionsetup{justification=justified, singlelinecheck=off} 
\caption*{\footnotesize \textit{Notes:} Event-study estimates across outcomes for ``experienced'' GitHub user accounts (created before 2017). The sample period spans March 27--30 (\textit{Pre}) and April 3--6 (\textit{Post}). All specifications include user, time, and day-of-the-week fixed effects. A log plus one transformation is applied to \textit{Avg. lines added per PR} (opened or merged). 95\% (90\%) confidence intervals for robust standard errors clustered at the user level are depicted in light (dark) grey.}
\end{figure}
\end{landscape}

\begin{landscape}
\begin{figure}[H]
\centering 
\caption{Pkg. Contributors -- Event-Study Estimates}
\label{fig:event-study-pkg}
\includegraphics[width=1.5\textwidth]{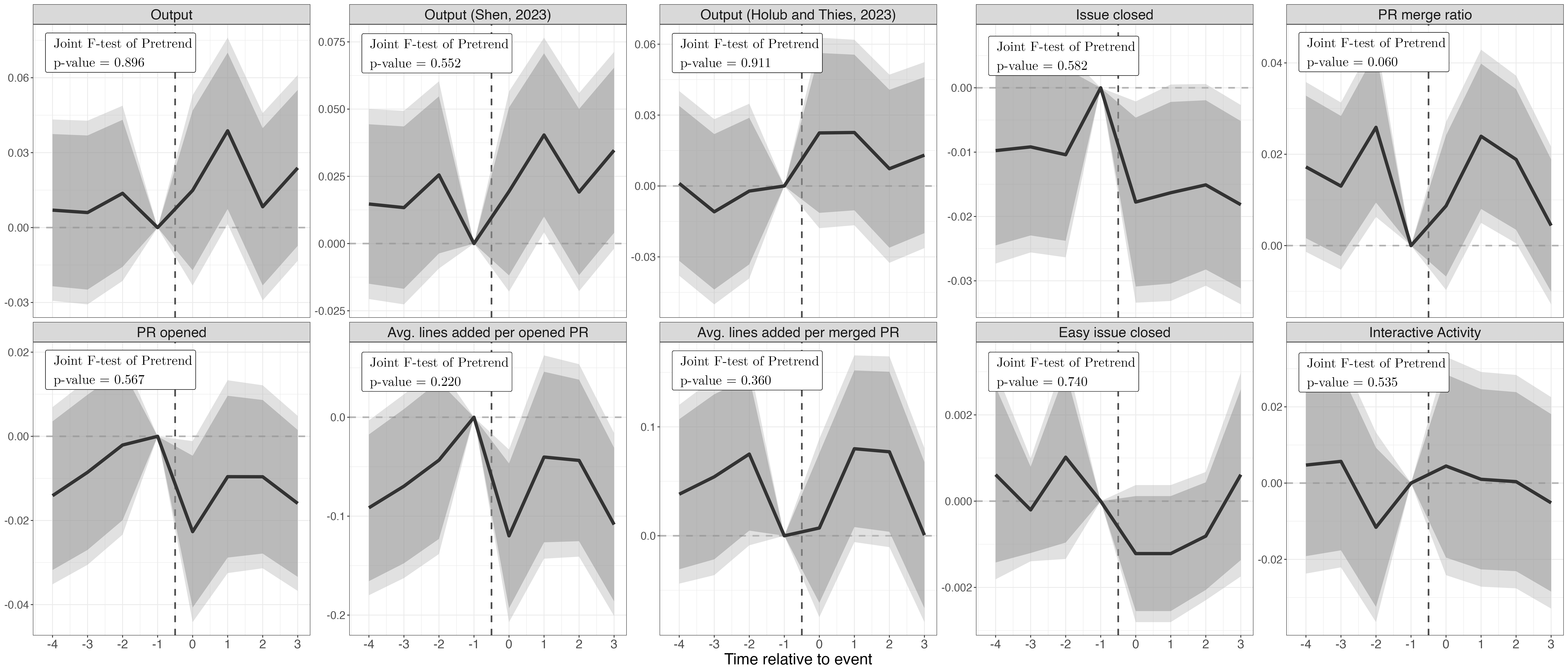}
\captionsetup{justification=justified, singlelinecheck=off} 
\caption*{\footnotesize \textit{Notes:} Event-study estimates across outcomes for ``pkg. contributor'' GitHub user accounts (owner and/or contributor to a programming package repository). The sample period spans March 27--30 (\textit{Pre}) and April 3--6 (\textit{Post}). All specifications include user, time, and day-of-the-week fixed effects. A log plus one transformation is applied to \textit{Avg. lines added per PR} (opened or merged). 95\% (90\%) confidence intervals for robust standard errors clustered at the user level are depicted in light (dark) grey.}
\end{figure}
\end{landscape}

\begin{table}[H]

\caption{Alternative Outcomes \label{tab:did-alt-out}}
\centering
\resizebox{\linewidth}{!}{
\begin{threeparttable}
\begin{tabular}[t]{lllccccc}
\toprule
 &  &  & (1) & (2) & (3) & (4) & (5)\\
\midrule
\addlinespace[0.3em]
\multicolumn{8}{l}{\textbf{A: Additional Outcomes}}\\
\hspace{1em} &  & \makecell{ \\ \\ } & \makecell{Any Event \\ \\ } & \makecell{Commit \\ \\} & \makecell{PR merged \\ \\ } & \makecell{PR merge \\ ratio (Holub \& \\ Thies, 2023)} & \makecell{Avg. files \\  edited per \\ merged PR}\\
\cmidrule{2-8}
\hspace{1em} & Overall & Treated $\times$ Post & 0.0091 & 0.0084 & 0.0047 & 0.0017 & 0.0119**\\

\hspace{1em} & (N = 290,864) &  & (0.0090) & (0.0075) & (0.0031) & (0.0018) & (0.0060)\\
\cmidrule(l{3pt}r{3pt}){3-8}

\hspace{1em} &  & Dep. var. mean & 0.3358 & 0.2215 & 0.0371 & 0.0130 & 0.0630\\
\cmidrule{2-8}
\hspace{1em} & Less experienced & Treated $\times$ Post & 0.0282** & 0.0220** & 0.0118*** & 0.0037 & 0.0224***\\

\hspace{1em} & (N = 149,680) &  & (0.0124) & (0.0106) & (0.0035) & (0.0024) & (0.0074)\\
\cmidrule(l{3pt}r{3pt}){3-8}

\hspace{1em} &  & Dep. var. mean & 0.3191 & 0.2272 & 0.0265 & 0.0122 & 0.0498\\
\cmidrule{2-8}
\hspace{1em} & Experienced & Treated $\times$ Post & -0.0116 & -0.0061 & -0.0029 & -0.0004 & 0.0005\\

\hspace{1em} & (N = 141,184) &  & (0.0133) & (0.0105) & (0.0053) & (0.0028) & (0.0098)\\
\cmidrule(l{3pt}r{3pt}){3-8}

\hspace{1em} &  & Dep. var. mean & 0.3534 & 0.2155 & 0.0483 & 0.0138 & 0.0769\\
\cmidrule{2-8}
\hspace{1em} & Pkg. contributor & Treated $\times$ Post & 0.0311 & 0.0242 & 0.0107 & -0.0024 & 0.0277\\

\hspace{1em} & (N = 47,328) &  & (0.0233) & (0.0199) & (0.0107) & (0.0050) & (0.0195)\\
\cmidrule(l{3pt}r{3pt}){3-8}

\hspace{1em} &  & Dep. var. mean & 0.4088 & 0.2521 & 0.0635 & 0.0171 & 0.0971\\
\cmidrule{1-8}
\addlinespace[0.3em]
\multicolumn{8}{l}{\textbf{B: Intensive Margin}}\\
\hspace{1em} &  & \makecell{ \\ \\} & \makecell{Outputs \\ \\} & \makecell{Outputs \\ (Shen, 2023) \\ } & \makecell{Outputs \\ (Holub \& \\  Thies, 2023)} & \makecell{PRs merged \\  \\ } & \makecell{PRs opened \\ \\ }\\
\cmidrule{2-8}
\hspace{1em} & Overall & Treated $\times$ Post & 0.0122 & 0.0164 & 0.0180 & 0.0053* & 0.0010\\

\hspace{1em} & (N = 290,864) &  & (0.0145) & (0.0146) & (0.0149) & (0.0032) & (0.0030)\\
\cmidrule(l{3pt}r{3pt}){3-8}

\hspace{1em} &  & Dep. var. mean & 0.3706 & 0.3727 & 0.4215 & 0.0347 & 0.0345\\
\cmidrule{2-8}
\hspace{1em} & Less experienced & Treated $\times$ Post & 0.0407** & 0.0464** & 0.0478** & 0.0144*** & 0.0041\\

\hspace{1em} & (N = 149,680) &  & (0.0190) & (0.0191) & (0.0193) & (0.0036) & (0.0035)\\
\cmidrule(l{3pt}r{3pt}){3-8}

\hspace{1em} &  & Dep. var. mean & 0.3538 & 0.3545 & 0.3830 & 0.0247 & 0.0276\\
\cmidrule{2-8}
\hspace{1em} & Experienced & Treated $\times$ Post & -0.0191 & -0.0164 & -0.0144 & -0.0045 & -0.0025\\

\hspace{1em} & (N = 141,184) &  & (0.0222) & (0.0224) & (0.0231) & (0.0053) & (0.0049)\\
\cmidrule(l{3pt}r{3pt}){3-8}

\hspace{1em} &  & Dep. var. mean & 0.3884 & 0.3919 & 0.4624 & 0.0454 & 0.0418\\
\cmidrule{2-8}
\hspace{1em} & Pkg. contributor & Treated $\times$ Post & -0.0002 & 0.0033 & 0.0313 & 0.0108 & -0.0166*\\

\hspace{1em} & (N = 47,328) &  & (0.0445) & (0.0445) & (0.0454) & (0.0108) & (0.0101)\\
\cmidrule(l{3pt}r{3pt}){3-8}

\hspace{1em} &  & Dep. var. mean & 0.4802 & 0.4852 & 0.5865 & 0.0589 & 0.0556\\
\bottomrule
\end{tabular}
\begin{tablenotes}[para]
\item \textit{Notes:} 
\item All specifications include user--fixed effects, day-of-the-week--fixed effects, and a linear time trend for control and treatment group. The ``Less experienced'' sample includes all Github user accounts created after or in 2017 (median), while the ``Experienced'' sample comprises all GitHub user accounts created before 2017. The ``Pkg. contributor'' sample comprises all GitHub user accounts that are the owner and/or contributor to a (analytical) programming package repository. The number of observations is depicted in parentheses after each sample definition. A log plus one transformation is applied to \textit{Avg. files edited per PR} and all intensive margin outcomes. Robust standard errors in parentheses are clustered on the user-level: * $p<0.1$, ** $p<0.05$, *** $p<0.01$.
\end{tablenotes}
\end{threeparttable}}
\end{table}

\begin{table}[H]

\caption{Placebo Effect in Calendar Week prior to ChatGPT Ban \label{tab:did-placebo-2023}}
\centering
\resizebox{\linewidth}{!}{
\begin{threeparttable}
\begin{tabular}[t]{lllccccc}
\toprule
 &  &  & (1) & (2) & (3) & (4) & (5)\\
\midrule
\addlinespace[0.3em]
\multicolumn{8}{l}{\textbf{A: Output Quantity and Quality}}\\
\hspace{1em} &  & \makecell{ \\ \\ } & \makecell{Output \\ \\ } & \makecell{Output \\ (Shen, 2023) \\} & \makecell{Output \\ (Holub \& \\ Thies, 2023)} & \makecell{Issue \\ closed \\} & \makecell{PR merge \\ ratio \\ }\\
\cmidrule{2-8}
\hspace{1em} & Overall & Treated $\times$ Post & -0.0010 & -0.0014 & -0.0052 & -0.0016 & -0.0070**\\

\hspace{1em} & (N = 290,864) &  & (0.0070) & (0.0069) & (0.0072) & (0.0022) & (0.0029)\\
\cmidrule(l{3pt}r{3pt}){3-8}

\hspace{1em} &  & Dep. var. mean & 0.2164 & 0.2138 & 0.2444 & 0.0136 & 0.0347\\
\cmidrule{2-8}
\hspace{1em} & Less experienced & Treated $\times$ Post & -0.0031 & -0.0028 & -0.0038 & -0.0021 & -0.0061*\\

\hspace{1em} & (N = 149,680) &  & (0.0097) & (0.0096) & (0.0099) & (0.0023) & (0.0033)\\
\cmidrule(l{3pt}r{3pt}){3-8}

\hspace{1em} &  & Dep. var. mean & 0.2163 & 0.2144 & 0.2323 & 0.0083 & 0.0249\\
\cmidrule{2-8}
\hspace{1em} & Experienced & Treated $\times$ Post & 0.0004 & -0.0010 & -0.0078 & -0.0010 & -0.0076\\

\hspace{1em} & (N = 141,184) &  & (0.0100) & (0.0099) & (0.0106) & (0.0038) & (0.0049)\\
\cmidrule(l{3pt}r{3pt}){3-8}

\hspace{1em} &  & Dep. var. mean & 0.2164 & 0.2131 & 0.2573 & 0.0192 & 0.0452\\
\cmidrule{2-8}
\hspace{1em} & Pkg. contributor & Treated $\times$ Post & 0.0096 & 0.0094 & 0.0000 & -0.0012 & 0.0020\\

\hspace{1em} & (N = 47,328) &  & (0.0189) & (0.0185) & (0.0198) & (0.0082) & (0.0099)\\
\cmidrule(l{3pt}r{3pt}){3-8}

\hspace{1em} &  & Dep. var. mean & 0.2627 & 0.2582 & 0.3208 & 0.0271 & 0.0604\\
\cmidrule{1-8}
\addlinespace[0.3em]
\multicolumn{8}{l}{\textbf{B: Task Choice and Complexity}}\\
\hspace{1em} &  & \makecell{ \\ \\} & \makecell{PR opened \\ \\ } & \makecell{Avg. lines \\  added per \\ opened PR} & \makecell{Avg. lines \\ added per \\ merged PR} & \makecell{Easy issue \\ closed \\} & \makecell{Interactive \\ Activity \\}\\
\cmidrule{2-8}
\hspace{1em} & Overall & Treated $\times$ Post & -0.0043 & -0.0226 & -0.0206 & -0.0006 & -0.0044\\

\hspace{1em} & (N = 290,864) &  & (0.0032) & (0.0142) & (0.0131) & (0.0005) & (0.0039)\\
\cmidrule(l{3pt}r{3pt}){3-8}

\hspace{1em} &  & Dep. var. mean & 0.0381 & 0.1509 & 0.1383 & 0.0006 & 0.0606\\
\cmidrule{2-8}
\hspace{1em} & Less experienced & Treated $\times$ Post & -0.0069* & -0.0246 & -0.0308* & -0.0006 & -0.0027\\

\hspace{1em} &  (N = 149,680) &  & (0.0037) & (0.0178) & (0.0157) & (0.0005) & (0.0042)\\
\cmidrule(l{3pt}r{3pt}){3-8}

\hspace{1em} &  & Dep. var. mean & 0.0300 & 0.1285 & 0.1087 & 0.0004 & 0.0315\\
\cmidrule{2-8}
\hspace{1em} & Experienced & Treated $\times$ Post & -0.0015 & -0.0213 & -0.0079 & -0.0007 & -0.0062\\

\hspace{1em} & (N = 141,184) &  & (0.0053) & (0.0228) & (0.0217) & (0.0009) & (0.0069)\\
\cmidrule(l{3pt}r{3pt}){3-8}

\hspace{1em} &  & Dep. var. mean & 0.0467 & 0.1747 & 0.1696 & 0.0009 & 0.0913\\
\cmidrule{2-8}
\hspace{1em} & Pkg. contributor & Treated $\times$ Post & -0.0018 & -0.0397 & 0.0249 & -0.0021 & 0.0002\\

\hspace{1em} & (N = 47,328) &  & (0.0107) & (0.0438) & (0.0427) & (0.0018) & (0.0149)\\
\cmidrule(l{3pt}r{3pt}){3-8}

\hspace{1em} &  & Dep. var. mean & 0.0656 & 0.2366 & 0.2193 & 0.0012 & 0.1376\\
\bottomrule
\end{tabular}
\begin{tablenotes}[para]
\item \textit{Notes:} 
\item The \textit{placebo} post-treatment period ranges from Mon 27 March 2023 to Thu 30 March 2023; the corresponding pre-treatment period is from Mon 20 March 2023 to Thu 23 March 2023. All specifications include user--fixed effects, day-of-the-week--fixed effects, and a linear time trend for control and treatment group. The ``Less experienced'' sample includes all Github user accounts created after or in 2017 (median), while the ``Experienced'' sample comprises all accounts created before 2017. The ``Pkg. contributor'' sample comprises all GitHub user accounts that are the owner and/or contributor to a package repository. The number of observations is depicted in parentheses after each sample definition. A log plus one transformation is applied to \textit{Avg. lines added per PR} (opened or merged). Robust standard errors in parentheses are clustered on the user-level: * $p<0.1$, ** $p<0.05$, *** $p<0.01$.
\end{tablenotes}
\end{threeparttable}}
\end{table}

\begin{table}[H]

\caption{Placebo Effect in Calendar Week prior to Easter 2022 \label{tab:did-placebo-2022}}
\centering
\resizebox{\linewidth}{!}{
\begin{threeparttable}
\begin{tabular}[t]{lllccccc}
\toprule
 &  &  & (1) & (2) & (3) & (4) & (5)\\
\midrule
\addlinespace[0.3em]
\multicolumn{8}{l}{\textbf{A: Output Quantity and Quality}}\\
\hspace{1em} &  & \makecell{ \\ \\ } & \makecell{Output \\ \\ } & \makecell{Output \\ (Shen, 2023) \\} & \makecell{Output \\ (Holub \& \\ Thies, 2023} & \makecell{Issue \\ closed \\} & \makecell{PR merge \\ ratio \\ }\\
\cmidrule{2-8}
\hspace{1em} & Overall & Treated $\times$ Post & 0.0109 & 0.0100 & 0.0083 & -0.0008 & 0.0034\\

\hspace{1em} & (N = 145,480) &  & (0.0110) & (0.0109) & (0.0115) & (0.0037) & (0.0049)\\
\cmidrule(l{3pt}r{3pt}){3-8}

\hspace{1em} &  & Dep. var. mean & 0.2492 & 0.2455 & 0.2872 & 0.0196 & 0.0424\\
\cmidrule{2-8}
\hspace{1em} & Less experienced & Treated $\times$ Post & 0.0159 & 0.0152 & 0.0152 & -0.0051 & 0.0054\\

\hspace{1em} & (N = 58,400) &  & (0.0174) & (0.0174) & (0.0179) & (0.0045) & (0.0061)\\
\cmidrule(l{3pt}r{3pt}){3-8}

\hspace{1em} &  & Dep. var. mean & 0.2448 & 0.2424 & 0.2669 & 0.0114 & 0.0274\\
\cmidrule{2-8}
\hspace{1em} & Experienced & Treated $\times$ Post & 0.0079 & 0.0069 & 0.0037 & 0.0024 & 0.0023\\

\hspace{1em} & (N = 87,080) &  & (0.0142) & (0.0141) & (0.0149) & (0.0054) & (0.0071)\\
\cmidrule(l{3pt}r{3pt}){3-8}

\hspace{1em} &  & Dep. var. mean & 0.2522 & 0.2475 & 0.3008 & 0.0252 & 0.0524\\
\cmidrule{2-8}
\hspace{1em} & Pkg. contributor & Treated $\times$ Post & 0.0326 & 0.0261 & 0.0225 & 0.0077 & -0.0015\\

\hspace{1em} & (N = 35,680) &  & (0.0243) & (0.0240) & (0.0252) & (0.0103) & (0.0127)\\
\cmidrule(l{3pt}r{3pt}){3-8}

\hspace{1em} &  & Dep. var. mean & 0.2853 & 0.2800 & 0.3456 & 0.0311 & 0.0609\\
\cmidrule{1-8}
\addlinespace[0.3em]
\multicolumn{8}{l}{\textbf{B: Task Choice and Complexity}}\\
\hspace{1em} &  & & \makecell{PR opened \\ \\ } & \makecell{Avg. lines \\  added per \\ opened PR} & \makecell{Avg. lines \\ added per \\ merged PR} & \makecell{Easy issue \\ closed \\} & \makecell{Interactive \\ Activity \\}\\
\cmidrule{2-8}
\hspace{1em} & Overall & Treated $\times$ Post & 0.0021 & 0.0240 & 0.0413* & 0.0000 & -0.0071\\

\hspace{1em} &  (N = 145,480) &  & (0.0050) & (0.0232) & (0.0214) & (0.0005) & (0.0066)\\
\cmidrule(l{3pt}r{3pt}){3-8}

\hspace{1em} &  & Dep. var. mean & 0.0446 & 0.1718 & 0.1593 & 0.0008 & 0.0845\\
\cmidrule{2-8}
\hspace{1em} & Less experienced & Treated $\times$ Post & 0.0028 & 0.0171 & 0.0354 & -0.0003 & -0.0037\\

\hspace{1em} & (N = 58,400) &  & (0.0067) & (0.0338) & (0.0271) & (0.0005) & (0.0081)\\
\cmidrule(l{3pt}r{3pt}){3-8}

\hspace{1em} &  & Dep. var. mean & 0.0330 & 0.1382 & 0.1117 & 0.0006 & 0.0441\\
\cmidrule{2-8}
\hspace{1em} & Experienced & Treated $\times$ Post & 0.0018 & 0.0297 & 0.0472 & 0.0002 & -0.0095\\

\hspace{1em} & (N = 87,080) &  & (0.0070) & (0.0315) & (0.0310) & (0.0008) & (0.0096)\\
\cmidrule(l{3pt}r{3pt}){3-8}

\hspace{1em} &  & Dep. var. mean & 0.0523 & 0.1943 & 0.1913 & 0.0010 & 0.1116\\
\cmidrule{2-8}
\hspace{1em} & Pkg. contributor & Treated $\times$ Post & 0.0006 & 0.0170 & 0.0462 & -0.0010 & 0.0037\\

\hspace{1em} &  (N = 35,680) &  & (0.0128) & (0.0572) & (0.0493) & (0.0012) & (0.0178)\\
\cmidrule(l{3pt}r{3pt}){3-8}

\hspace{1em} &  & Dep. var. mean & 0.0665 & 0.2376 & 0.2159 & 0.0014 & 0.1443\\
\bottomrule
\end{tabular}
\begin{tablenotes}[para]
\item \textit{Notes:} 
\item The \textit{placebo} post-treatment period ranges from Mon 11 April 2022 to Thu 14 April 2022; the corresponding pre-treatment period is from Mon 4 April 2022 to Thu 7 April 2022. All specifications include user--fixed effects, day-of-the-week--fixed effects, and a linear time trend for control and treatment group. The ``Less experienced'' sample includes all Github user accounts created after or in 2017 (median), while the ``Experienced'' sample comprises all accounts created before 2017. The ``Pkg. contributor'' sample comprises all GitHub user accounts that are the owner and/or contributor to package repository. The number of observations is depicted in parentheses after each sample definition. A log plus one transformation is applied to \textit{Avg. lines added per PR} (opened or merged). Robust standard errors in parentheses are clustered on the user-level: * $p<0.1$, ** $p<0.05$, *** $p<0.01$.
\end{tablenotes}
\end{threeparttable}}
\end{table}

\begin{table}[H]

\caption{Effect of ChatGPT Ban on Working Activity \label{tab:did-work-hours}}
\centering
\resizebox{0.9\linewidth}{!}{
\begin{threeparttable}
\begin{tabular}[t]{lllccccc}
\toprule
 &  &  & (1) & (2) & (3) & (4) & (5)\\
\midrule
\addlinespace[0.3em]
\multicolumn{8}{l}{\textbf{A: Any Activity}}\\
\hspace{1em} &  & \makecell{ \\ \\ } & \makecell{``Work'' \\ hours\\ } & \makecell{First \\ activity \\} & \makecell{Last \\ activity \\ } & \makecell{Unusual \\ activity \\ (10th \%tile)} & \makecell{Unusual \\ activity \\ (25th \%tile)}\\
\cmidrule{2-8}
\hspace{1em} & Overall & Treated $\times$ Post & 0.0356 & -0.2048 & -0.1933 & 0.0032 & 0.0053\\

\hspace{1em} &  (N = 290,864) &  & (0.0497) & (0.1777) & (0.1784) & (0.0038) & (0.0063)\\
\cmidrule(l{3pt}r{3pt}){3-8}

\hspace{1em} &  & Dep. var. mean & 0.9620 & 11.3925 & 14.5319 & 0.0494 & 0.1292\\
\cmidrule{2-8}
\hspace{1em} & Less experienced & Treated $\times$ Post & 0.1344** & -0.2833 & -0.0351 & 0.0083 & 0.0130\\

\hspace{1em} & (N = 149,680) &  & (0.0643) & (0.2543) & (0.2506) & (0.0052) & (0.0084)\\
\cmidrule(l{3pt}r{3pt}){3-8}

\hspace{1em} &  & Dep. var. mean & 0.8479 & 11.7579 & 14.6761 & 0.0477 & 0.1196\\
\cmidrule{2-8}
\hspace{1em} & Experienced & Treated $\times$ Post & -0.0742 & -0.1427 & -0.3592 & -0.0026 & -0.0029\\

\hspace{1em} & (N = 141,184) &  & (0.0772) & (0.2489) & (0.2543) & (0.0055) & (0.0094)\\
\cmidrule(l{3pt}r{3pt}){3-8}

\hspace{1em} &  & Dep. var. mean & 1.0829 & 11.0537 & 14.3982 & 0.0512 & 0.1395\\
\cmidrule{2-8}
\hspace{1em} & Pkg. contributor & Treated $\times$ Post & 0.1684 & -0.0484 & -0.1477 & 0.0069 & 0.0226\\

\hspace{1em} & (N = 47,328) &  & (0.1535) & (0.3616) & (0.3987) & (0.0107) & (0.0180)\\
\cmidrule(l{3pt}r{3pt}){3-8}

\hspace{1em} &  & Dep. var. mean & 1.4032 & 10.8529 & 14.4870 & 0.0584 & 0.1652\\
\cmidrule{1-8}
\addlinespace[0.3em]
\multicolumn{8}{l}{\textbf{B: Productive Activities}}\\
\hspace{1em} &  & \makecell{ \\ \\} & \makecell{``Work'' \\ hours\\ } & \makecell{First \\ activity \\} & \makecell{Last \\ activity \\ } & \makecell{Unusual \\ activity \\ (10th \%tile)} & \makecell{Unusual \\ activity \\ (25th \%tile)}\\
\cmidrule{2-8}
\hspace{1em} & Overall & Treated $\times$ Post & 0.0346 & 0.0159 & 0.0083 & -0.0004 & -0.0010\\

\hspace{1em} & (N = 290,864) &  & (0.0428) & (0.2032) & (0.2017) & (0.0032) & (0.0053)\\
\cmidrule(l{3pt}r{3pt}){3-8}

\hspace{1em} &  & Dep. var. mean & 0.7395 & 11.4108 & 14.6531 & 0.0374 & 0.0955\\
\cmidrule{2-8}
\hspace{1em} & Less experienced & Treated $\times$ Post & 0.1200** & -0.2090 & -0.0059 & 0.0031 & 0.0082\\

\hspace{1em} &  (N = 149,680) &  & (0.0567) & (0.2848) & (0.2782) & (0.0045) & (0.0073)\\
\cmidrule(l{3pt}r{3pt}){3-8}

\hspace{1em} &  & Dep. var. mean & 0.6884 & 11.7987 & 14.7975 & 0.0383 & 0.0946\\
\cmidrule{2-8}
\hspace{1em} & Experienced & Treated $\times$ Post & -0.0591 & 0.2347 & 0.0107 & -0.0043 & -0.0110\\

\hspace{1em} & (N = 141,184) &  & (0.0651) & (0.2906) & (0.2937) & (0.0045) & (0.0077)\\
\cmidrule(l{3pt}r{3pt}){3-8}

\hspace{1em} &  & Dep. var. mean & 0.7937 & 11.0049 & 14.5019 & 0.0365 & 0.0965\\
\cmidrule{2-8}
\hspace{1em} & Pkg. contributor & Treated $\times$ Post & 0.0822 & 0.0431 & -0.2626 & -0.0041 & 0.0075\\

\hspace{1em} &  (N = 47,328) &  & (0.1296) & (0.4233) & (0.4664) & (0.0086) & (0.0150)\\
\cmidrule(l{3pt}r{3pt}){3-8}

\hspace{1em} &  & Dep. var. mean & 0.9935 & 10.8366 & 14.4380 & 0.0416 & 0.1147\\
\bottomrule
\end{tabular}
\begin{tablenotes}[para]
\item \textit{Notes:} 
\item All specifications include user--fixed effects, day-of-the-week--fixed effects, and a linear time trend for control and treatment group. ``Productive activities'' comprise: \texttt{PullRequestEvent},  \texttt{PullRequestReviewEvent}, \texttt{PullRequestReviewCommentEvent}, \texttt{PushEvent}, \texttt{ReleaseEvent}, \texttt{CreateEvent}, \texttt{IssueEvent}. The number of observations is depicted in parentheses after each sample definition. For \emph{first} and \emph{last activity} only users with at least one activity on the day are included; the number of observations for the ``overall'', ``less experienced'', ``experienced'', and ``pkg. contributor'' sample in Panel A, respectively, Panel B are 43,868, 40,672, 84,540, 17,802 and 30,717, 32,135, 62,852, 12,663. Robust standard errors in parentheses are clustered on the user-level: * $p<0.1$, ** $p<0.05$, *** $p<0.01$.
\end{tablenotes}
\end{threeparttable}}
\end{table}

\begin{figure}[H]
\centering 
\caption{``Leave-One-Out'' Analysis}
\label{fig:leave-one-out}
\includegraphics[width=\textwidth]{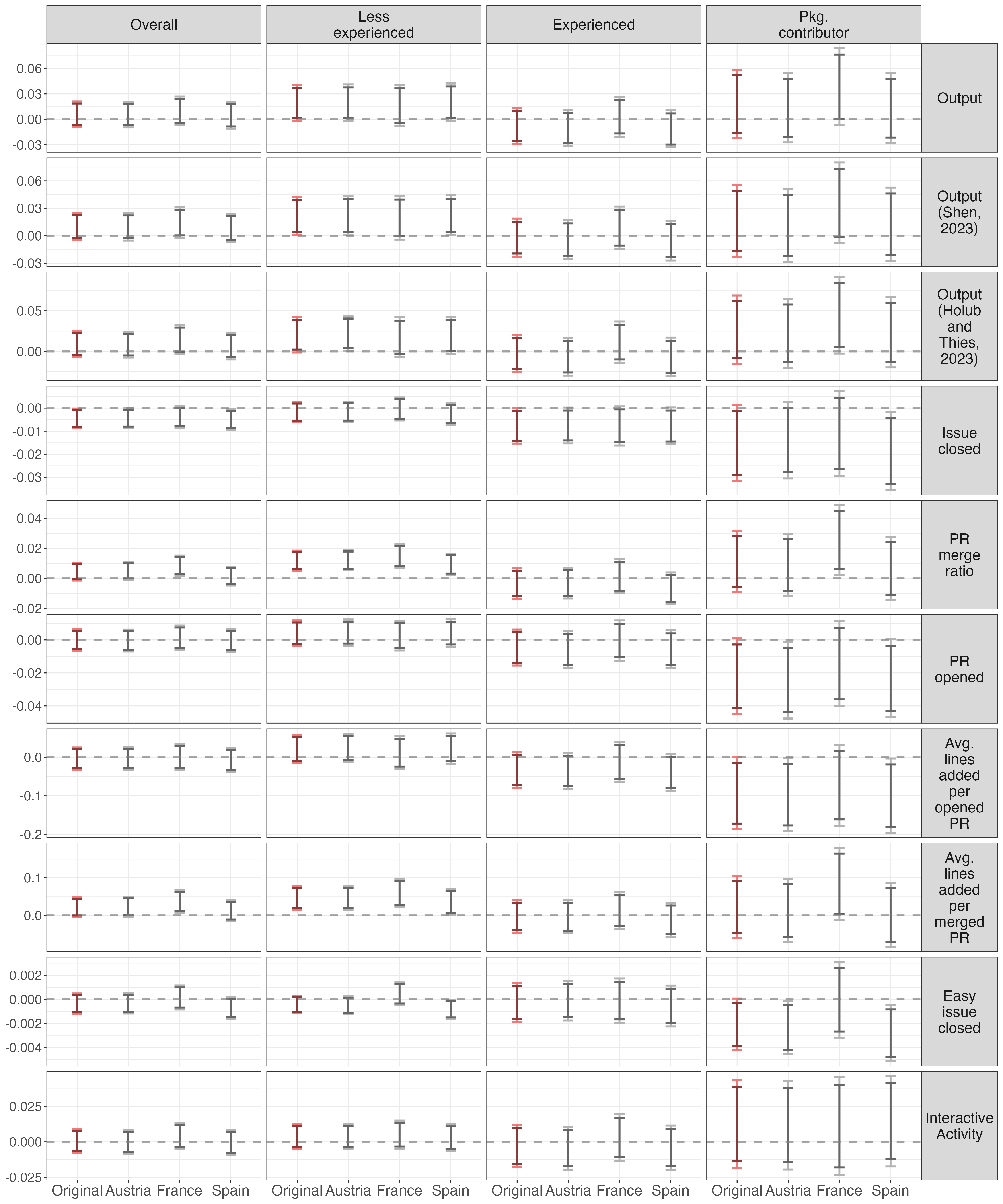}
\captionsetup{justification=justified, singlelinecheck=off} 
\caption*{\footnotesize \textit{Notes:} Treatment effect estimates for the baseline DID specification are presented when users from one of the control group countries---i.e. Austria, France or Spain---are consecutively dropped from the sample. Point estimates and 95\% (90\%) cluster-robust confidence intervals are depicted in light (dark) grey for ``Leave-One-Out'' estimates and juxtaposed to the ``Original'' DID estimates (c. Table \ref{tab:did-main}) in light (dark) red.}
\end{figure}

\begin{table}[H]

\caption{Effect of ChatGPT Ban accounting for Multiple Hypothesis Testing \label{tab:did-rwolf}}
\centering
\resizebox{\linewidth}{!}{
\begin{threeparttable}
\begin{tabular}[t]{lllccccc}
\toprule
 &  &  & (1) & (2) & (3) & (4) & (5)\\
\midrule
\addlinespace[0.3em]
\multicolumn{8}{l}{\textbf{A: Output Quantity and Quality}}\\
\hspace{1em} &  & \makecell{ \\ \\ } & \makecell{Output \\ \\ } & \makecell{Output \\ (Shen, 2023) \\} & \makecell{Output \\ (Holub \& \\ Thies, 2023} & \makecell{Issue \\ closed \\} & \makecell{PR merge \\ ratio \\ }\\
\cmidrule{2-8}
\hspace{1em} & Overall (N = 290,864) & Treated $\times$ Post & 0.0062 & 0.0102 & 0.0090 & -0.0045** & 0.0045\\

\hspace{1em} &  &  & {}[0.960] & {}[0.602] & {}[0.823] & {}[0.083] & {}[0.426]\\
\cmidrule(l{3pt}r{3pt}){3-8}

\hspace{1em} &  & Dep. var. mean & 0.2314 & 0.2283 & 0.2624 & 0.0142 & 0.0359\\
\cmidrule{2-8}
\hspace{1em} & Less experienced (N = 149,680) & Treated $\times$ Post & 0.0193* & 0.0216** & 0.0203* & -0.0017 & 0.0118***\\

\hspace{1em} &  &  & {}[0.179] & {}[0.085] & {}[0.157] & {}[0.960] & {}[0.000]\\
\cmidrule(l{3pt}r{3pt}){3-8}

\hspace{1em} &  & Dep. var. mean & 0.2339 & 0.2316 & 0.2519 & 0.0089 & 0.0258\\
\cmidrule{2-8}
\hspace{1em} & Experienced (N = 141,184) & Treated $\times$ Post & -0.0079 & -0.0020 & -0.0030 & -0.0077* & -0.0034\\

\hspace{1em} &  &  & {}[0.960] & {}[0.999] & {}[0.998] & {}[0.110] & {}[0.971]\\
\cmidrule(l{3pt}r{3pt}){3-8}

\hspace{1em} &  & Dep. var. mean & 0.2287 & 0.2249 & 0.2734 & 0.0198 & 0.0467\\
\cmidrule{2-8}
\hspace{1em} & Pkg. contributor (N = 47,328) & Treated $\times$ Post & 0.0181 & 0.0164 & 0.0270 & -0.0151* & 0.0113\\

\hspace{1em} &  &  & {}[0.945] & {}[0.960] & {}[0.710] & {}[0.179] & {}[0.835]\\
\cmidrule(l{3pt}r{3pt}){3-8}

\hspace{1em} &  & Dep. var. mean & 0.2688 & 0.2644 & 0.3312 & 0.0267 & 0.0615\\
\cmidrule{1-8}
\addlinespace[0.3em]
\multicolumn{8}{l}{\textbf{B: Task Choice and Complexity}}\\
\hspace{1em} &  & \makecell{ \\ \\} & \makecell{PR opened \\ \\ } & \makecell{Avg. lines \\  added per \\ opened PR} & \makecell{Avg. lines \\ added per \\ merged PR} & \makecell{Easy issue \\ closed \\} & \makecell{Interactive \\ Activity \\}\\
\cmidrule{2-8}
\hspace{1em} & Overall (N = 290,864) & Treated $\times$ Post & 0.0000 & -0.0039 & 0.0220 & -0.0004 & 0.0006\\

\hspace{1em} &  &  & {}[0.999] & {}[0.998] & {}[0.311] & {}[0.956] & {}[0.999]\\
\cmidrule(l{3pt}r{3pt}){3-8}

\hspace{1em} &  & Dep. var. mean & 0.0402 & 0.1581 & 0.1439 & 0.0007 & 0.0637\\
\cmidrule{2-8}
\hspace{1em} & Less experienced (N = 149,680) & Treated $\times$ Post & 0.0040 & 0.0211 & 0.0456*** & -0.0004 & 0.0037\\

\hspace{1em} &  &  & {}[0.898] & {}[0.823] & {}[0.004] & {}[0.823] & {}[0.960]\\
\cmidrule(l{3pt}r{3pt}){3-8}

\hspace{1em} &  & Dep. var. mean & 0.0315 & 0.1358 & 0.1148 & 0.0005 & 0.0340\\
\cmidrule{2-8}
\hspace{1em} & Experienced (N = 141,184) & Treated $\times$ Post & -0.0046 & -0.0324 & -0.0031 & -0.0003 & -0.0028\\

\hspace{1em} &  &  & {}[0.960] & {}[0.581] & {}[0.999] & {}[0.996] & {}[0.996]\\
\cmidrule(l{3pt}r{3pt}){3-8}

\hspace{1em} &  & Dep. var. mean & 0.0494 & 0.1817 & 0.1749 & 0.0009 & 0.0953\\
\cmidrule{2-8}
\hspace{1em} & Pkg. contributor (N = 47,328) & Treated $\times$ Post & -0.0221* & -0.0933* & 0.0226 & -0.0021* & 0.0127\\

\hspace{1em} &  &  & {}[0.132] & {}[0.109] & {}[0.989] & {}[0.126] & {}[0.960]\\
\cmidrule(l{3pt}r{3pt}){3-8}

\hspace{1em} &  & Dep. var. mean & 0.0661 & 0.2356 & 0.2210 & 0.0014 & 0.1393\\
\bottomrule
\end{tabular}
\begin{tablenotes}[para]
\item \textit{Notes:} 
\item All specifications include user--fixed effects, day-of-the-week--fixed effects, and a linear time trend for the control and treatment group. The ``Less experienced'' sample includes all Github user accounts created after or in 2017 (median), while the ``Experienced'' sample comprises all GitHub user accounts created before 2017. The ``Pkg. contributor'' sample comprises all GitHub user accounts that are the owner and/or contributor to a (analytical) programming package repository. The number of observations is depicted in parentheses after each sample definition. A log plus one transformation is applied to \textit{Avg. lines added per PR} (opened or merged). The Romano–Wolf p-values adjusted for multiple hypothesis testing are presented in brackets and calculated with the resampled null distribution from 1000 bootstrap samples with the \textit{Stata} command \texttt{rwolf} \citepappx{clarke2020}. Robust standard errors are clustered on the user-level: * $p<0.1$, ** $p<0.05$, *** $p<0.01$.
\end{tablenotes}
\end{threeparttable}}
\end{table}

\section{GitHub Repository--User-Level Data}\label{sec:repo-results}

\begin{table}[H]

\caption{Descriptive Statistics \label{tab:summary-user}}
\centering
\resizebox{\linewidth}{!}{
\begin{threeparttable}
\begin{tabular}[t]{lrrrrrrrr}
\toprule
\multicolumn{1}{c}{ } & \multicolumn{2}{c}{Overall} & \multicolumn{2}{c}{Less experienced} & \multicolumn{2}{c}{Experienced} & \multicolumn{2}{c}{Pkg. contributor} \\
\multicolumn{1}{c}{ } & \multicolumn{2}{c}{(N = 4,627 $ \times $ 11,938)} & \multicolumn{2}{c}{(N = 3,315 $ \times $ 4,566)} & \multicolumn{2}{c}{(N = 4,294 $ \times $ 7,372)} & \multicolumn{2}{c}{(N = 3,006 $ \times $ 2,902)} \\
\cmidrule(l{3pt}r{3pt}){2-3} \cmidrule(l{3pt}r{3pt}){4-5} \cmidrule(l{3pt}r{3pt}){6-7} \cmidrule(l{3pt}r{3pt}){8-9}
  & Mean & SD & Mean & SD & Mean & SD & Mean & SD\\
\midrule
\addlinespace[0.3em]
\multicolumn{9}{l}{\textbf{A - Repo-User-Day--Level (N = 239,912)}}\\
\hline
\hspace{1em}Output & 0.0116 & 0.1069 & 0.0080 & 0.0889 & 0.0136 & 0.1158 & 0.0171 & 0.1296\\
\hspace{1em}Output (Shen, 2023) & 0.0108 & 0.1036 & 0.0075 & 0.0860 & 0.0128 & 0.1122 & 0.0163 & 0.1264\\
\hspace{1em}Output (Holub and Thies, 2023) & 0.0279 & 0.1647 & 0.0192 & 0.1373 & 0.0327 & 0.1780 & 0.0405 & 0.1971\\
\hspace{1em}Issue closed & 0.0024 & 0.0492 & 0.0014 & 0.0376 & 0.0030 & 0.0546 & 0.0034 & 0.0580\\
\hspace{1em}PR merge ratio & 0.0030 & 0.0542 & 0.0013 & 0.0358 & 0.0039 & 0.0621 & 0.0047 & 0.0679\\
\hspace{1em}PR opened & 0.0054 & 0.0734 & 0.0041 & 0.0641 & 0.0061 & 0.0780 & 0.0080 & 0.0890\\
\hspace{1em}Avg. lines added per opened PR & 0.0170 & 0.2762 & 0.0134 & 0.2494 & 0.0190 & 0.2902 & 0.0238 & 0.3190\\
\hspace{1em}Avg. lines added per merged PR & 0.0104 & 0.2167 & 0.0047 & 0.1483 & 0.0135 & 0.2469 & 0.0153 & 0.2541\\
\hspace{1em}Easy issue closed & 0.0001 & 0.0094 & 0.0000 & 0.0034 & 0.0001 & 0.0114 & 0.0001 & 0.0089\\
\hspace{1em}Interactive Activity & 0.0193 & 0.1377 & 0.0127 & 0.1120 & 0.0230 & 0.1500 & 0.0285 & 0.1663\\
\addlinespace[0.3em]
\hline
\multicolumn{9}{l}{\textbf{B - User--Level (N = 11,938)}}\\
\hline
\hspace{1em} & Mean & SD & Min & Median & Max &  &  & \\
\cmidrule(l{3pt}r{3pt}){1-6}
\hspace{1em}User creation year & 2015.37 & 3.51 & 2009 & 2015 & 2023 &  &  & \\
\hspace{1em}Experienced & 0.62 & 0.49 & 0 & 1 & 1 &  &  & \\
\hspace{1em}Pkg. contributions & 78.06 & 563.84 & 0 & 0 & 19514 &  &  & \\
\hspace{1em}Pkg. owner & 0.07 & 0.26 & 0 & 0 & 1 &  &  & \\
\hspace{1em}Followers & 49.65 & 300.24 & 0 & 12 & 17421 &  &  & \\
\hspace{1em}Following & 36.54 & 313.69 & 0 & 11 & 28300 &  &  & \\
\hspace{1em}Repositories & 40.40 & 78.31 & 0 & 23 & 3900 &  &  & \\
\hspace{1em}Total events & 2.89 & 7.47 & 0 & 1 & 98 &  &  & \\
\bottomrule
\end{tabular}
\begin{tablenotes}[para]
\item \textit{Notes:} 
\item Panel A presents descriptive statistics for the baseline sample period Pre 27-30.03 -- Post 03-06.04. The ``Less experienced'' sample includes all GitHub user accounts created after or in 2017 (median), while the ``Experienced'' sample comprises all GitHub user accounts created before 2017. The ``Pkg. contributor'' sample comprises all GitHub user accounts that are the owner and/or contributor to a (analytical) programming package repository. The number of unique GitHub user accounts for the entire repository $\times$ user sample (``Overall'') and each of the subsamples is presented in parentheses below. A log plus one transformation is applied to \textit{Avg. lines added per PR} (opened or merged). Panel B provides information on the individual characteristics of all GitHub user accounts in the repository $\times$ user sample.
\end{tablenotes}
\end{threeparttable}}
\end{table}

\clearpage

\begin{table}[H]

\caption{Effect of ChatGPT Ban on GitHub Output at the Repository-Level \label{tab:did-repo-main}}
\centering
\resizebox{\linewidth}{!}{
\begin{threeparttable}
\begin{tabular}[t]{lllccccc}
\toprule
 &  &  & (1) & (2) & (3) & (4) & (5)\\
\midrule
\addlinespace[0.3em]
\multicolumn{8}{l}{\textbf{A: Output Quantity and Quality}}\\
\hspace{1em} &  & \makecell{ \\ \\ } & \makecell{Output \\ \\ } & \makecell{Output \\ (Shen, 2023) \\} & \makecell{Output \\ (Holub \& \\ Thies, 2023} & \makecell{Issue \\ closed \\} & \makecell{PR merge \\ ratio \\ }\\
\cmidrule{2-8}
\hspace{1em} & Overall & Treated $\times$ Post & -0.0010 & -0.0007 & 0.0029* & -0.0007 & -0.0002\\

\hspace{1em} & (N = 239,912) &  & (0.0010) & (0.0010) & (0.0016) & (0.0005) & (0.0005)\\
\cmidrule(l{3pt}r{3pt}){3-8}

\hspace{1em} &  & Dep. var. mean & 0.0116 & 0.0108 & 0.0279 & 0.0024 & 0.0030\\
\cmidrule{2-8}
\hspace{1em} & Less experienced & Treated $\times$ Post & 0.0019 & 0.0014 & 0.0047** & 0.0007 & 0.0014***\\

\hspace{1em} & (N = 86,200) &  & (0.0015) & (0.0014) & (0.0022) & (0.0006) & (0.0005)\\
\cmidrule(l{3pt}r{3pt}){3-8}

\hspace{1em} &  & Dep. var. mean & 0.0080 & 0.0075 & 0.0192 & 0.0014 & 0.0013\\
\cmidrule{2-8}
\hspace{1em} & Experienced & Treated $\times$ Post & -0.0027** & -0.0019 & 0.0018 & -0.0016** & -0.0011\\

\hspace{1em} & (N = 153,712) &  & (0.0014) & (0.0013) & (0.0022) & (0.0007) & (0.0007)\\
\cmidrule(l{3pt}r{3pt}){3-8}

\hspace{1em} &  & Dep. var. mean & 0.0136 & 0.0128 & 0.0327 & 0.0030 & 0.0039\\
\cmidrule{2-8}
\hspace{1em} & Pkg. contributor & Treated $\times$ Post & -0.0036 & -0.0021 & 0.0011 & -0.0025** & 0.0001\\

\hspace{1em} & (N = 63,064) &  & (0.0025) & (0.0024) & (0.0036) & (0.0012) & (0.0011)\\
\cmidrule(l{3pt}r{3pt}){3-8}

\hspace{1em} &  & Dep. var. mean & 0.0171 & 0.0163 & 0.0405 & 0.0034 & 0.0047\\
\cmidrule{1-8}
\addlinespace[0.3em]
\multicolumn{8}{l}{\textbf{B: Task Choice and Complexity}}\\
\hspace{1em} &  & \makecell{ \\ \\} & \makecell{PR opened \\ \\ } & \makecell{Avg. lines \\  added per \\ opened PR} & \makecell{Avg. lines \\ added per \\ merged PR} & \makecell{Easy issue \\ closed \\} & \makecell{Interactive \\ Activity \\}\\
\cmidrule{2-8}
\hspace{1em} & Overall & Treated $\times$ Post & 0.0008 & 0.0022 & -0.0008 & 0.0001 & 0.0027**\\

\hspace{1em} & (N = 239,912) &  & (0.0007) & (0.0027) & (0.0020) & (0.0001) & (0.0013)\\
\cmidrule(l{3pt}r{3pt}){3-8}

\hspace{1em} &  & Dep. var. mean & 0.0054 & 0.0170 & 0.0104 & 0.0001 & 0.0193\\
\cmidrule{2-8}
\hspace{1em} & Less experienced & Treated $\times$ Post & 0.0019* & 0.0059 & 0.0073*** & 0.0000 & 0.0028\\

\hspace{1em} & (N = 86,200) &  & (0.0011) & (0.0044) & (0.0024) & (0.0000) & (0.0018)\\
\cmidrule(l{3pt}r{3pt}){3-8}

\hspace{1em} &  & Dep. var. mean & 0.0041 & 0.0134 & 0.0047 & 0.0000 & 0.0127\\
\cmidrule{2-8}
\hspace{1em} & Experienced & Treated $\times$ Post & 0.0001 & 0.0002 & -0.0058** & 0.0001 & 0.0026\\

\hspace{1em} & (N = 153,712) &  & (0.0010) & (0.0034) & (0.0029) & (0.0002) & (0.0018)\\
\cmidrule(l{3pt}r{3pt}){3-8}

\hspace{1em} &  & Dep. var. mean & 0.0061 & 0.0190 & 0.0135 & 0.0001 & 0.0230\\
\cmidrule{2-8}
\hspace{1em} & Pkg. contributor & Treated $\times$ Post & 0.0003 & -0.0030 & -0.0015 & -0.0002 & 0.0031\\

\hspace{1em} & (N = 63,064) &  & (0.0016) & (0.0058) & (0.0043) & (0.0002) & (0.0031)\\
\cmidrule(l{3pt}r{3pt}){3-8}

\hspace{1em} &  & Dep. var. mean & 0.0080 & 0.0238 & 0.0153 & 0.0001 & 0.0285\\
\bottomrule
\end{tabular}
\begin{tablenotes}[para]
\item \textit{Notes:} 
\item All specifications include repository $\times$ user fixed effects and day-of-the-week--fixed effects. The ``Less experienced'' sample includes all Github user accounts created after or in 2017 (median), while the ``Experienced'' sample comprises all GitHub user accounts created before 2017. The ``Pkg. contributor'' sample comprises all GitHub user accounts that are the owner and/or contributor to a (analytical) programming package repository. The number of observations is depicted in parentheses after each sample definition. A log plus one transformation is applied to \textit{Avg. lines added per PR} (opened or merged). Robust standard errors in parentheses are clustered on the repository $times$ user-level: * $p<0.1$, ** $p<0.05$, *** $p<0.01$.
\end{tablenotes}
\end{threeparttable}}
\end{table}

\begin{table}[H]

\caption{Effect of ChatGPT Ban on \texttt{PyPI} Repository Contributors' Output \label{tab:did-repo-lang}}
\centering
\resizebox{\linewidth}{!}{
\begin{threeparttable}
\begin{tabular}[t]{llccccc}
\toprule
 &  & (1) & (2) & (3) & (4) & (5)\\
\midrule
 & \makecell{ \\ \\ } & \makecell{Output \\ \\ } & \makecell{Output \\ (Shen, 2023) \\} & \makecell{Output \\ (Holub \& \\ Thies, 2023} & \makecell{Issue \\ closed \\} & \makecell{PR merge \\ ratio \\ }\\
\cmidrule{1-7}
 & Treated $\times$ Post & -0.0042** & -0.0022 & -0.0007 & -0.0021** & -0.0011\\

 &  & (0.0021) & (0.0021) & (0.0040) & (0.0009) & (0.0007)\\
\cmidrule(l{3pt}r{3pt}){2-7}

\multirow{-3}{*}{\raggedright\arraybackslash Overall (N = 33,512)} & Dep. var. mean & 0.0078 & 0.0073 & 0.0227 & 0.0016 & 0.0016\\
\cmidrule{1-7}
 & Treated $\times$ Post & -0.0004 & 0.0017 & 0.0027 & -0.0019* & 0.0004\\

 &  & (0.0029) & (0.0028) & (0.0057) & (0.0010) & (0.0007)\\
\cmidrule(l{3pt}r{3pt}){2-7}

\multirow{-3}{*}{\raggedright\arraybackslash Less experienced (N = 12,784)} & Dep. var. mean & 0.0055 & 0.0052 & 0.0173 & 0.0007 & 0.0003\\
\cmidrule{1-7}
 & Treated $\times$ Post & -0.0069** & -0.0048* & -0.0031 & -0.0023* & -0.0021*\\

 &  & (0.0029) & (0.0029) & (0.0056) & (0.0013) & (0.0011)\\
\cmidrule(l{3pt}r{3pt}){2-7}

\multirow{-3}{*}{\raggedright\arraybackslash Experienced (N = 20,728)} & Dep. var. mean & 0.0093 & 0.0086 & 0.0261 & 0.0022 & 0.0023\\
\cmidrule{1-7}
 & Treated $\times$ Post & -0.0060 & -0.0024 & 0.0057 & -0.0037** & -0.0025\\

 &  & (0.0042) & (0.0042) & (0.0082) & (0.0019) & (0.0017)\\
\cmidrule(l{3pt}r{3pt}){2-7}

\multirow{-3}{*}{\raggedright\arraybackslash Pkg. contributor (N = 11,392)} & Dep. var. mean & 0.0121 & 0.0123 & 0.0364 & 0.0011 & 0.0020\\
\bottomrule
\end{tabular}
\begin{tablenotes}[para]
\item \textit{Notes:} 
\item All specifications include repository $\times$ user fixed effects and day-of-the-week--fixed effects. The ``Less experienced'' sample includes all Github user accounts created after or in 2017 (median), while the ``Experienced'' sample comprises all GitHub user accounts created before 2017. The ``Pkg. contributor'' sample comprises all GitHub user accounts that are the owner and/or contributor to a (analytical) programming package repository. The number of observations is depicted in parentheses after each sample definition. A log plus one transformation is applied to \textit{Avg. lines added per PR} (opened or merged). Robust standard errors in parentheses are clustered on the repository $times$ user-level: * $p<0.1$, ** $p<0.05$, *** $p<0.01$.
\end{tablenotes}
\end{threeparttable}}
\end{table}

% \input{tables/did-table-appx-2728-0304-repo}

% \input{tables/did-table-appx-2730-0306-repo}

% \begin{figure}[H]
% \centering 
% \caption{Repository-Level Analysis}
% \label{fig:event-study-repo}
% \includegraphics[width=\textwidth]{figures/event-study-repo-baseline.png}
% \captionsetup{justification=justified, singlelinecheck=off} 
% \caption*{\small \textit{Notes:} Panel A displays event-study estimates across outcomes for ``less experienced'' GitHub user accounts (created after or in 2017). Panel B presents event-study estimates for \textit{(i)} ``experienced'' GitHub user accounts (created before 2017) and \textit{(ii)} and accounts that are the owner and/or contributor to a(n) (analytical) programming package repository (``Pkg. contributor''). The sample period spans March 27--30 (\textit{Pre}) and April 3--6 (\textit{Post}). All specifications include user $\times$ repository, time, and day-of-the-week fixed effects; 95\% (90\%) confidence intervals for robust standard errors clustered at the user $\times$ repository level are depicted in light (dark) grey.}
% \end{figure}

\section{User Adaption to the ChatGPT Ban}\label{sec:google-tor}

Considering that there appears to be mean reversion in the estimated effects toward the end of our sample period, we now turn our attention to adaptation behaviour. The simplest way to circumvent the ChatGPT ban was to use VPN tools or encrypted routing through, for instance, the TOR network.

\subsection{Data} 
\begin{figure}[!h]
\centering 
\caption{Panel Structure of Google Trends and TOR Data}
\label{fig:panelView}
\includegraphics[width=\textwidth]{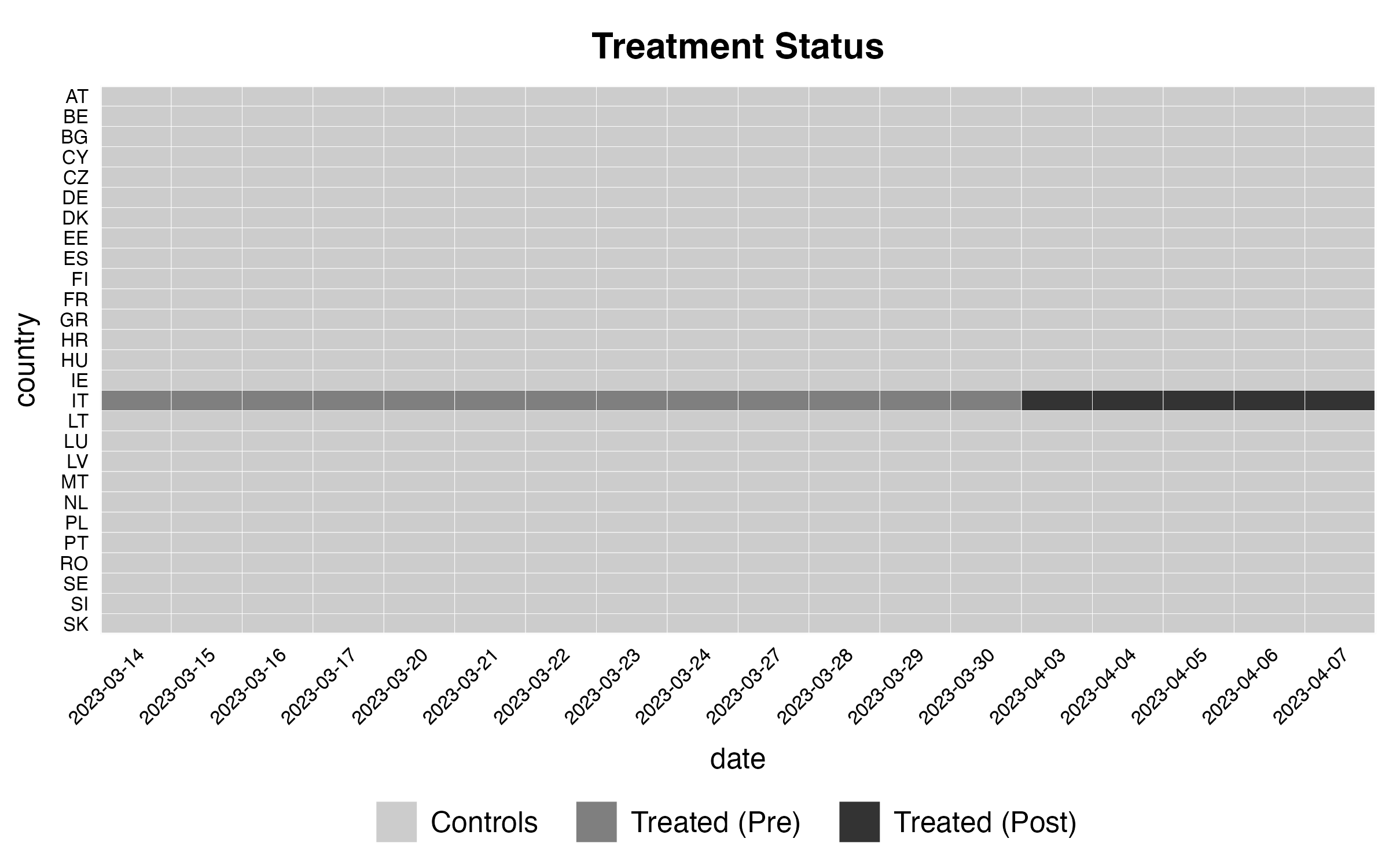}
\captionsetup{justification=justified, singlelinecheck=off} 
\caption*{\small \textit{Notes:} The panel structure of the datasets used in Section \ref{sec:google-tor}---i.e., the \textit{(i) Google trends} and the \textit{(ii) TOR} (\textit{``standard''} and \textit{bridge relay}) user datasets---are displayed. Workday dates during the sample period are presented on the x-axis, and \textit{ISO 3166-1 alpha-2} country codes are presented on the y-axis.}
\end{figure}

We collect daily data on the number of \textit{Google searches} on the topic of ``Virtual Private Networks'' from \textit{Google Trends} and on the number of users of \textit{TOR}, an open-source software for enabling anonymous communication, from \textit{TOR Metrics} for all 25 countries in the European Union.\footnote{Google trends data have been widely used in economic research as a predictor of human behavioural economic phenomena \citepappx{choi2012}. For example, \citetappx{bohme2020} used Google trends data on migration-related Google search terms to predict international migration, while \citetappx{ginsberg2009} used trends data to predict influenza outbreaks.} We retrieve information on both the number of users of ``standard'' \textit{TOR relays} and of \textit{TOR bridge} relays to examine whether there were changes in the use of, in particular, \textit{TOR bridge} relays, which are not listed publicly and therefore are more difficult for firewalls to identify.\footnote{Note that \textit{TOR bridge} relays can, however, slow down the connection. For more information on bridges vs. ``standard'' relays, please refer to the official \textit{TOR} documentation at \url{https://tb-manual.torproject.org/bridges/}.} We apply a log transformation to both user numbers. The sample period under consideration covers March 13, 2023, the day after the release of ChatGPT-4, until April 7, 2023, the end of the workweek post-ban. Observations on weekends are dropped from the sample since we are interested in the effect of the ban on output. Figure \ref{fig:panelView} provides a graphic illustration of the final panel structure.

\subsection{Results} To estimate the average treatment effect of the ChatGPT ban on users from Italy, we apply the generalized synthetic control method proposed by \citetappx{xu2017}. The treatment effect on the treated unit (ATT) is the difference between the actual outcome and its estimated counterfactual. To obtain the counterfactual, a (cross-validated) interactive fixed effects (IFE) model is estimated for the control group data.\footnote{Specifically, we apply the EM algorithm proposed by \citetappx{Gobillon2016} and implemented in the \textbf{R} package \texttt{gsynth} \citepappx{gsynth2022}, which additionally uses treatment group information for the pre-treatment period, leading to (slightly) more precisely estimated coefficients.} All IFE models incorporate additive unit and time fixed effects.\footnote{Note that the \textit{Google trends} data are already standardized by country for the selected time period such that we include only time fixed effects in this case when estimating the IFE model.} To draw inference, we rely on the parametric bootstrap procedure suggested by \citetappx{xu2017} for settings with a small number of treated units. 

\begin{figure}[!h]
\centering 
\caption{Effect of ChatGPT Ban on Ban Circumvention Technology}
\label{fig:ban-circumvention}
\includegraphics[width=\textwidth]{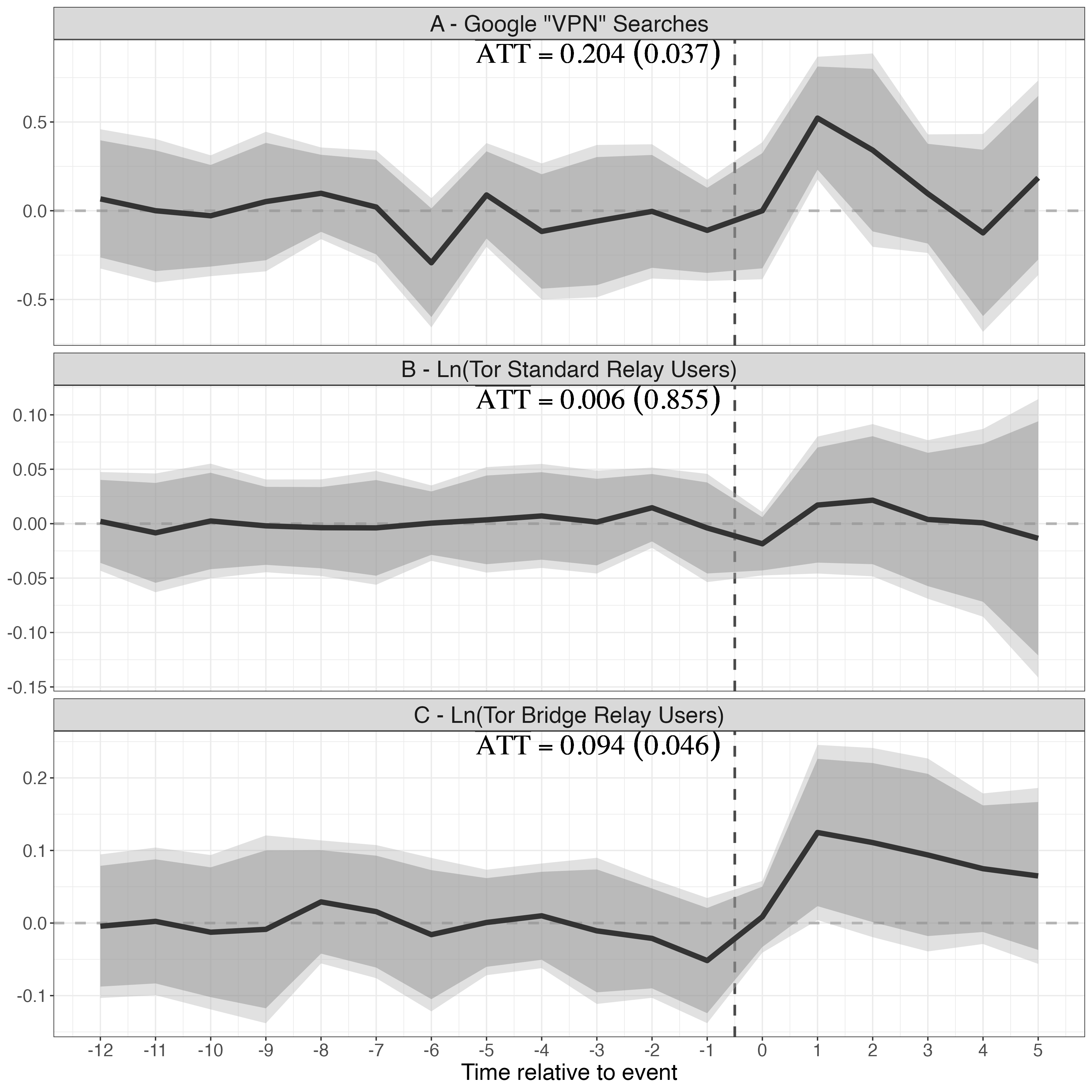}
\captionsetup{justification=justified, singlelinecheck=off} 
\caption*{\small \textit{Notes:} The dynamic treatment effects estimates for the generalized synthetic control method of \citetappx{xu2017} are depicted. The top panel presents the ATT for the number of \textit{Google queries} on the topic of VPNs. The bottom panel presents the ATT for \textit{TOR bridge} relay users. The counterfactual for the treated unit (Italy) is estimated with an interactive fixed effects model; 95\% (90\%) confidence intervals from the parametric bootstrap procedure proposed by \citetappx{xu2017} are displayed in light (dark) grey. Additionally, the \textit{mean} ATT over the workweek after the ChatGPT ban and its $p$-value (in parentheses) are presented.}
\end{figure}

The top panel in Figure \ref{fig:ban-circumvention} presents the effect of the ChatGPT ban on the number of \textit{Google queries} on the topic of VPNs. We observe a significant increase in the share of web searches on this relative to other topics in Italy on the first working day after the ban that slowly vanishes over the next three days. The estimated effect on April 3 is sizeable: the share of searches on VPNs increases by 52.2 percentage points. On average, the share of queries on VPNs was 20.6 percentage points higher in Italy over the workweek. The observed pattern is consistent with Italian users looking for ways to access ChatGPT even after the ban and succeeding after some initial search costs. Our estimates might, however, present only \textit{stated} preferences. 

To investigate whether the ban actually led to behavioral changes among Italian users, we look at an alternative outcome: the log number of \textit{TOR} users. The results for \textit{TOR relay} and \textit{TOR bridge} users are presented in Panels B and C of Figure \ref{fig:ban-circumvention}, respectively. While the number of \textit{TOR relay} users shows only a minor increase in the days after the ban, the average treatment effect on the number of Italian \textit{TOR bridge} users is positive and significant on the first workday after the ban. Usage of \textit{TOR bridges} remained elevated for the entire workweek, with an increase in user numbers---on average---of approximately 9.4 percentage points. This pattern is in line with users resorting to \textit{bridge} over ``standard'' relays to minimize the chance of their being denied access to ChatGPT since the former are more difficult for firewalls to identify.\footnote{For a discussion on denial of  ChatGPT access, see the following OpenAI forum discussion: \url{https://community.openai.com/t/access-denied-error-1020/38758/23}.}

Overall, our findings are consistent with Italian users looking for and finding ways to circumvent the blocked access to ChatGPT.\\

\bibliographystyleappx{aer}
\bibliographyappx{bibliography.bib}

\end{document}